\documentclass[draftclsnofoot, onecolumn, 12pt]{IEEEtran}

\usepackage{hyperref} % create hyperlinks
\usepackage{graphicx}
\usepackage{amssymb}
\usepackage{amsmath}
\usepackage{mathtools}
\usepackage{cite}
\usepackage{stfloats}
\usepackage{subfigure}
\usepackage{epstopdf}
\usepackage{dsfont}
\usepackage{psfrag}
\usepackage[mathscr]{euscript}
\usepackage{acronym}  % make an acronym
\usepackage{booktabs} 
\usepackage[table]{xcolor}
\usepackage{algorithm}
\usepackage{algpseudocode}
\usepackage{algcompatible}

\acrodef{CCDF}{complementary cumulative distribution function}
\acrodef{CF}{characteristic function}
\acrodef{PPP}{Poisson point process}
\acrodef{CSI}{channel state information}
\acrodef{OFDM}{orthogonal frequency division multiplexing}
\acrodef{OFDMA}{orthogonal frequency division multiple access}

\acrodef{RV}{random variable}
\acrodef{i.i.d.}{independent, identically distributed}
\acrodef{PMF}{probability mass function}
\acrodef{PDF}{probability distribution function}
\acrodef{CDF}{cumulative distribution function}
\acrodef{PGFL}{probability generating functional}
\acrodef{ch.f.}{characteristic function}
\acrodef{AWGN}{additive white Gaussian noise}
\acrodef{SNR}{signal-to-noise ratio}
\acrodef{LRT}{likelihood ratio test}
\acrodef{DRT}{distance ratio test}
\acrodef{GLRT}{generalized likelihood ratio test}
\acrodef{CRLB}{Cram\'{e}r-Rao lower bound}
\acrodef{CRB}{Cram\'{e}r-Rao bound}
\acrodef{ZZLB}{Ziv-Zakai lower bound}
\acrodef{ZZB}{Ziv-Zakai bound}
\acrodef{LoS}{line-of-sight}
\acrodef{ToF}{time-of-flight}
\acrodef{NLoS}{non-line-of-sight}
\acrodef{GDOP}{geometric dilution of precision}
\acrodef{GPS}{Global Positioning System}
\acrodef{FIM}{Fisher information matrix}
\acrodef{PEB}{position error bound}
\acrodef{SPEB}{squared position error bound}
\acrodef{TOA}{time-of-arrival}
\acrodef{TOF}{time-of-flight}
\acrodef{WSN}{wireless sensor network}
\acrodef{MAC}{medium access control}
\acrodef{RSS}{received signal strength}
\acrodef{WAF}{wall attenuation factor}
\acrodef{TDOA}{time difference-of-arrival}
\acrodef{RF}{radiofrequency}
\acrodef{RTT}{round-trip time}
\acrodef{AOA}{angle-of-arrival}
\acrodef{MF}{matched filter}
\acrodef{ED}{energy detector}
\acrodef{ML}{maximum likelihood}
\acrodef{MSE}{mean-square error}
\acrodef{RMSE}{root-mean-square error}
\acrodef{LEO}{localization error outage}
\acrodef{ppm}{part-per-million}
\acrodef{ACK}{acknowledge}
\acrodef{NACK}{negative-acknowledge}
\acrodef{UWB}{Ultrawide bandwidth}
\acrodef{TNR}{threshold-to-noise ratio}
\acrodef{LS}{least squares}
\acrodef{IR-UWB}{impulse radio UWB}
\acrodef{FCC}{Federal Communications Commission}
\acrodef{TH}{time-hopping}
\acrodef{PPM}{pulse position modulation}
\acrodef{MUI}{multi-user interference}
\acrodef{PDP}{power delay profile}
\acrodef{BPZF}{band-pass zonal filter}
\acrodef{SIR}{signal-to-interference ratio}
\acrodef{RFID}{radio frequency identification}
\acrodef{WPAN}{wireless personal area network}
\acrodef{WWB}{Weiss-Weinstein bound}
\acrodef{DP}{direct path}
\acrodef{MF}{matched filter}
\acrodef{MMSE}{minimum-mean-square-error}
\acrodef{SBS}{serial backward search}
\acrodef{SBSMC}{serial backward search for multiple clusters}
\acrodef{NBI}{narrowband interference}
\acrodef{WBI}{wideband interference}
\acrodef{INR}{interference-to-noise ratio}
\acrodef{CR}{channel response}
\acrodef{CIR}{channel impulse response}
\acrodef{CR}{channel  response}
\acrodef{RADAR}{radar}
\acrodef{MUR}{Multistatic radar}
\acrodef{JBSF}{jump back and search forward}
\acrodef{HDSA}{high-definition situation-aware}
\acrodef{RRC}{root raised cosine}
\acrodef{ST}{simple thresholding}
\acrodef{BTB}{Bellini-Tartara bound}
\acrodef{P-Max}{$P$-Max}  %suggestion, use with \acl{P-Max}
\acrodef{MIMO}{multiple-input multiple-output}
\acrodef{MAP}{maximum a posteriori}
\acrodef{FG}{factor graph}
\acrodef{OP}{outage probability}
\acrodef{WED}{wall extra delay}
\acrodef{RMS}{root mean square}
\acrodef{SPAWN}{sum-product algorithm over a wireless network}
\acrodef{MDD}{minimum distance distribution}
\acrodef{MAP}{maximum a posteriori probability}
\acrodef{PAR}{probabilistic association rule}
\acrodef{AP}{access point}
\acrodef{HD}{half-duplex}
\acrodef{FD}{full-duplex}
\acrodef{IC}{interference cancellation}
\acrodef{HDHN}{hybrid-duplex heterogeneous network}
\acrodef{TDD}{time-division duplexing}
\acrodef{FDD}{frequency-division duplexing}
\acrodef{SINR}{signal-to-interference-plus-noise ratio}
\acrodef{UAV}{unmanned aerial vehicle}
\acrodef{GCS}{ground control station}
\acrodef{LTE}{long term evolution}
\acrodef{AoI}{age of information}
\acrodef{KKT}{Karush-Kuhn-Tucker}
\acrodef{IR-HARQ}{incremental redundancy hybrid automatic repeat request}
\acrodef{HARQ}{hybrid automatic repeat request}
\acrodef{AoA}{angle-of-arrival}
\acrodef{URLLC}{ultra-reliable low-latency communication}
\acrodef{MRT}{maximum ratio transmission}
\acrodef{RZF}{regularzied zero-forcing}
\acrodef{E2E}{end-to-end}
\acrodef{5G}{fifth generation}
\acrodef{ZF}{zero-forcing}
\acrodef{CSIT}{channel state information at transmitter}
\acrodef{QoS}{quality of service}
\acrodef{MRC}{maximum ratio combining}
\acrodef{GPI}{generalized power iteration}
\acrodef{SDP}{semi-definite programming}
\acrodef{DCTU}{delay-constrained and delay-tolerant user}
\acrodef{UHD}{ultra-high demension}
\acrodef{FEC}{forward error correction}
\acrodef{ARQ}{automatic repeat request}
\usepackage{color}
\usepackage{dsfont}
\usepackage{bbm}

\newcommand{\Ws}[2]{{W_{}^{}}} % Symbol bandwidth
\newcommand{\TSIR}[2]{{\tau_{}^{}}}

\DeclareMathAlphabet{\mathsf}{OML}{cmbr}{m}{it}

\newtheorem{theorem}{Theorem}
\newtheorem{lemma}{Lemma}

\newtheorem{remark}{Remark}

\newcommand{\bd}{\begin{description}}
\newcommand{\ed}{\end{description}}
\newcommand{\be}{\begin{enumerate}}
\newcommand{\ee}{\end{enumerate}}
\newcommand{\bi}{\begin{itemize}}
\newcommand{\ei}{\end{itemize}}
\newcommand{\bl}{\begin{list}}
\newcommand{\el}{\end{list}}
\newcommand{\bt}{\begin{tabbing}}
\newcommand{\et}{\end{tabbing}}

\acrodef{BS}{base station}
\acrodef{G2A}{Ground-to-Air}
\acrodef{A2G}{Air-to-Ground}
\acrodef{A2A}{Air-to-Air}
\acrodef{G2G}{Ground-to-Ground}
\acrodef{IoT}{Internet of things}
\acrodef{MU-MIMO}{multi-user multiple-input multiple-output}

\begin{document}

\newcommand{\paperTitle}{\huge Precoding Design for Multi-user MIMO Systems with Delay-Constrained and -Tolerant Users}
 
%---------------------------------------------------------------------------%
%                     title, title footnote, header                         %
%---------------------------------------------------------------------------%

% paper title
\title{\paperTitle}
% \title{Distributed Secrecy in \\Multilevel Wireless Networks}

% author names, IEEE memberships, corresponding address, title footnote %
\author{
	\IEEEauthorblockN{
		Minsu Kim, 
		Jeonghun Park, and 
		Jemin~Lee
	}\\[0.5em]
\thanks{
	M.\ Kim and J.\ Lee are with the Department of Information and
	Communication Engineering, Daegu Gyeongbuk Institute of Science and
	Technology, Daegu 42988, South Korea
	(e-mail: \texttt{ads5577@dgist.ac.kr}, \texttt{jmnlee@dgist.ac.kr}).
	
	J.\ Park is with the School of Electronics Engineering, Kyungpook National University, Daegu, 41566, South Korea
	(e-mail: \texttt{jeonghun.park@knu.ac.kr}).
}
\thanks{The material in this paper was presented, in part, at the International Conference on Communications, Montreal, Canada, Jun. 2021 \cite{Kim:21}
}
%	       \thanks{
%	       This work was supported in part by the the National Research Foundation of Korea (NRF) grant funded by the Korea government (MSIP) (No. 2017R1C1B2009280) and the DGIST R\&D Program of the Ministry of Science and ICT(17-ST-02).
%	       }
\thanks{
	The corresponding author is J. Lee. 
}
}

%% make the title area
%% Don't write page number 0 to the cover page.
\maketitle %% make the title area

%
% \markboth{Submitted to IEEE Journal on Selected Areas in Communications}{\title}

%
%%%%%%%%% uncomment this section for a 2-column formt %%%%%%%
%%%%%%%%% [begin] %%%%%%%%
%\thispagestyle{empty}
%  \textcolor{blue}{\framebox{\textsf{\small{Today: \today}}}}\\

%
%\newpage
%%%%%%%%% [end] %%%%%%%%
\setcounter{page}{1}
\acresetall
%%---------------------------------------------------------------------------%
%%                           abstract and key words                          %
%%---------------------------------------------------------------------------%
\begin{abstract}
	In both academia and industry, \ac{MU-MIMO} techniques have shown enormous gains in spectral efficiency by exploiting spatial degrees of freedom.
	So far, an underlying assumption in most of the existing \ac{MU-MIMO} design has been that all the users use infinite blocklength, so that they can achieve the Shannon capacity.
	This setup, however, is not suitable considering delay-constrained users whose blocklength tends to be finite. 
	In this paper, we consider a heterogeneous setting in \ac{MU-MIMO} systems where delay-constrained users and delay-tolerant users coexist, called a DCTU-MIMO network.
	To maximize the sum spectral efficiency in this system, we present the spectral efficiency for delay-tolerant users and provide a lower bound of the spectral efficiency for delay-constrained users.
	We consider an optimization problem that maximizes the sum spectral efficiency of delay-tolerant users while satisfying the latency constraint of delay-constrained users, and propose a \ac{GPI} precoding algorithm that finds a principal precoding vector.
	Furthermore, we extend a DCTU-MIMO network to the multiple time slots scenario and propose a recursive generalized power iteration precoding algorithm.
	In simulation results, we prove proposed methods outperform baseline schemes and present the effect of network parameters on the ergodic sum spectral efficiency.
\end{abstract}
%
%\clearpage
\acresetall
%
%
%%---------------------------------------------------------------------------%
%%                           Sec: Introduction                               %
%%---------------------------------------------------------------------------%
%
\section{Introduction}
In \ac{5G} communications, as the demand of real-time applications (e.g., virtual reality, smart healthcare, and autonomous driving) increases, the \ac{URLLC} has been considered as the emerging application scenario \cite{She:17,Fen:19,Pop:19,Par:20,Muk:20}.
Especially, the \ac{URLLC} is required to have high reliability (e.g., more than 99.999\%), low \ac{E2E} latency (e.g., less than 1ms), and small packet size (e.g., 32 bytes) \cite{3GPP:TR:36.881:V14.0.0}.
To meet those extreme requirements, several wireless communication techniques are being actively studied; a \ac{MIMO} technique is one of them. 
As the \ac{MIMO} techniques have shown enormous performance gains in the previous generations, it is also being expected that the \ac{MIMO} has a crucial role to support the \ac{5G} \ac{URLLC}. 
One key requirement to reap the high spectral efficiency of \ac{MIMO} systems is a delicate design of linear precoding \cite{Spe:04,Sad:07,Liu:12}. In general, it is known that the sum spectral efficiency optimization with respect to linear precoding is non-convex; thereby finding a global optimal solution is infeasible. 
Further, incorporating user scheduling, a problem becomes NP-hard.

% Conventional MU-MIMO
For resolving this difficulty, in the past few decades, many researchers have focused on the design of the user scheduling and the precoding in a \ac{MU-MIMO} network.
The works in \cite{Spen:04,Dim:05,Yoo:06,Jor:10} maximized the sum spectral efficiency subject to the total power constraint using the \ac{ZF} and \ac{MRT} precoding, which are classic methods.
However, they only optimized the user scheduling for the two precoding methods, which are not optimal in general power regime.
In \cite{Son:08}, the beamforming, the power control, and the user scheduling were optimized to maximize the minimum weighted rate among all users.
In \cite{Kho:09}, the authors maximized the sum rate by adopting a decoupled approach for designing the quantized codebook, the precoder, and the scheduler.
Extending to the multi-cell \ac{MU-MIMO} network, the works in \cite{Yu:13,Wee:13} presented the optimal user scheduling and precoding vector to maximize the weighted sum spectral efficiency for the multi-cell network.
However, the aforementioned works in \cite{Son:08,Kho:09,Yu:13,Wee:13} fail to optimize the user scheduling and the precoding jointly.

To overcome these limitations, some works in \cite{Ban:20,Cho:20,Han:19} proposed the joint user scheduling and precoding design algorithm.
In \cite{Ban:20}, the optimal user scheduling and precoding were presented for three objectives: the weighted sum spectral efficiency maximization, the minimum \ac{SINR} maximization, and the power consumption minimization.
In \cite{Cho:20}, the user selection, the power allocation, and the precoding were optimized to maximize the weighted sum spectral efficiency with the imperfect \ac{CSIT} for the multi-cell network.
In \cite{Han:19}, the authors maximized the sum spectral efficiency of a cell-free massive \ac{MIMO} network by jointly identifying a set of cooperative \acp{AP}, the precoding for beamforming and compression, and the power control.
However, the works in \cite{Ban:20,Cho:20,Han:19} only considered delay-tolerant users whose blocklength was implicitly assumed to be infinite. 
Hence, the solutions, provided in those works, are not applicable to the \ac{MIMO} network with a delay-constrained user, which is generally operating in the finite blocklength regime.
The delay-constrained users are the ones required to receive the information within a target latency.
It is essential to consider those delay-constrained users in the network design for the increasing real-time applications.

% Finite blocklength regime
Recently, some works considering delay-constrained users have been actively studied, especially for the \ac{URLLC} scenarios that require the extremely low \ac{E2E} latency (e.g., less than 1ms).
Many researchers tried to reduce the communication duration by decreasing the packet size, and consequently, the approximated maximal achievable rate in the finite blocklength regime, different from conventional Shannon capacity, was presented \cite{Pol:09,Yan:14}.
In the view of the finite blocklength regime, the optimal resource allocation is studied according to network parameters in \cite{Sun:19,Hu:20}.
In \cite{Sun:19}, the global optimal resource allocation ensuring the decoding error probability and the transmission delay was investigated. 
%The power allocation was optimized to maximize the sum throughput under the \ac{QoS} constraint while considering different types of data arrivals in \cite{Hu:18}.
The \ac{QoS}-constrained throughput with \ac{IR-HARQ} was analyzed to control the transmit power allocation in \cite{Hu:20}.
Furthermore, a few works in \cite{Ren:20,Gha:20,Nas:20} analyzed the sum rate in the finite blocklength regime when the \ac{MU-MIMO} network is considered.
In \cite{Ren:20}, the optimal pilot and payload transmission power was presented for both \ac{MRC} and \ac{ZF} methods to maximize the weighted uplink data rate with the imperfect \ac{CSI}.
In \cite{Gha:20}, the optimal beamforming vector was presented to maximize the weighted sum rate subject to the \ac{QoS} requirement of each user.
In \cite{Nas:20}, the authors maximized the minimum rate among users for two cases: optimizing the beamforming vector and optimizing the power allocation with the \ac{RZF} beamforming.
However, most of those works did not consider the \ac{MU-MIMO} network (such as \cite{Pol:09,Yan:14,Sun:19,Hu:20}), and some of those works considered delay-constrained users only in the \ac{MU-MIMO} network (such as \cite{Ren:20,Gha:20,Nas:20}).
Since the \ac{BS} supports various services (e.g., autonomous driving and \ac{UHD} video) in current and future networks, we need to carefully design the \ac{MU-MIMO} network where delay-constrained users and delay-tolerant users coexist.

Therefore, in this work, we consider a general setup of the \ac{MU-MIMO} network with \acp{DCTU}, which we denote it as the DCTU-MIMO network. 
The \ac{BS}, equipped with multiple transmit antennas, simultaneously serves delay-tolerant users and delay-constrained users.
We also consider the scenario that the \ac{IR-HARQ} is adopted to enhance the communication reliability of users.
To the best of our knowledge, this is the first work that considers not only the delay-tolerant users but also the delay-constrained users in the \ac{MU-MIMO} network.
In the DCTU-MIMO network, we analyze the spectral efficiencies for delay-constrained users and delay-tolerant users, respectively.
We then formulate an optimization problem that maximizes the sum spectral efficiency of delay-tolerant users while satisfying the latency requirement of delay-constrained users. 
We finally propose a \ac{GPI} precoding algorithm, defined as the Delay-GPI, that provides a joint solution for the user scheduling and the precoding at the \ac{BS} of the optimization problem.
The main contributions of this paper can be summarized as follows.
\begin{itemize}
	\item We develop the novel optimization framework of the DCTU-MIMO network without and with the \ac{IR-HARQ} scheme in the presence of delay-constrained users as well as delay-tolerant users.
	Especially, we consider the finite blocklength coding for the delay-constrained users while the infinite blocklength coding is used for the delay-tolerant users.
	%We also introduce the DCTU-MIMO network with the \ac{IR-HARQ} scheme to provide high-quality services by leveraging the time diversity.}
	%
	\item We analyze the spectral efficiency for two types of users without and with the \ac{IR-HARQ} scheme.
	After deriving an upper bound on the channel dispersion for the interference channel without the \ac{IR-HARQ} scheme, we also provide the lower bound on the spectral efficiency of the delay-constrained user as a Rayleigh quotient form, which is more tractable in the optimization.
	\item We consider the optimization problem that maximizes the sum spectral efficiency of delay-tolerant users while satisfying the latency constraint of delay-constrained users.
	We provide the first-order optimality condition of this problem and present the generalized power iteration precoding algorithm to find a feasible principal precoding vector that also satisfies the first-order optimality condition.
	\item We finally show that the proposed algorithm outperforms baseline methods in terms of the ergodic sum spectral efficiency. Especially, the proposed algorithm improves the ergodic sum spectral efficiency by allocating minimal transmission power that meets the communication latency requirement to delay-constrained users.
\end{itemize}

The remainder of this paper is organized as follows.
We introduce a DCTU-MIMO network in Section \ref{sec:System} and
analyze the spectral efficiency of delay-tolerant users and delay-constrained users in Section~\ref{sec:SE}.
From the spectral efficiency analysis, for the DCTU-MIMO network without the \ac{IR-HARQ} scheme, we formulate a sum spectral efficiency maximization problem in Section \ref{sec:Algorithm}. 
Furthermore, we derive a first-order optimality condition and propose an algorithm to find a solution of this problem.
In Section \ref{sec:Algorithm_IR-HARQ},
we formulate an optimization problem of the DCTU-MIMO network with the \ac{IR-HARQ} scheme and provide a computationally efficient algorithm.
In Section \ref{sec:Numerical},
we evaluate the performance of the DCTU-MIMO network according to network parameters and 
compare the ergodic sum spectral efficiency of the proposed algorithm with that of baseline methods.
Finally, the conclusion is presented in Section \ref{sec:Conclusion}.

\textit{Notation}: The conjugate transpose of $\textbf{x}$ is denoted by $\textbf{x}^\text{H}$ and the inverse matrix of $\textbf{x}$ is denoted by $\textbf{x}^{-1}$. In addition, $\textbf{I}_N$ is the identity matrix with size $N \times N$ and $\text{diag}\left(\textbf{A}_1, \cdots, \textbf{A}_N\right) \in \mathbb{C}^{NK \times NK}$ is a block diagonal matrix where $\textbf{A}_k \in \mathbb{C}^{K \times K}$ is a square matrix.

%
%%---------------------------------------------------------------------------%
%%                           Sec: System Model                               %
%%---------------------------------------------------------------------------%
%
\section{System Model} \label{sec:System}
In this section, we first introduce a DCTU-MIMO network.
We then explain the \ac{IR-HARQ} scheme and describe the DCTU-MIMO network with \ac{IR-HARQ} scheme.

\subsection{Network Model without IR-HARQ scheme} \label{subsec:system_without}
\begin{figure}[t!]
	\begin{center}   
		{ 
			\includegraphics[width=0.55\columnwidth]{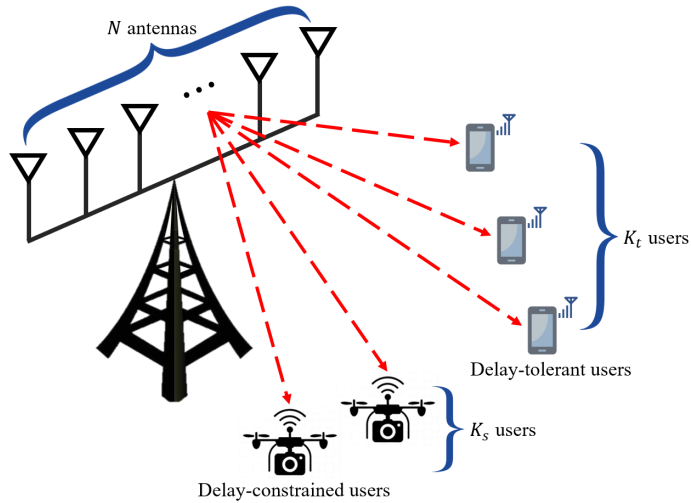}
			\vspace{-8mm}
		}
	\end{center}
	\caption{
		DCTU-MIMO network where a \ac{BS} equipped with multiple antennas simultaneously serves delay-tolerant users and delay-constrained users with a single antenna.
		\vspace{-3mm}
	}
	\label{fig:system}
\end{figure}

As shown in Fig.~\ref{fig:system}, we consider a DCTU-MIMO network, where a \ac{BS} equipped with $N$ antennas transmits the downlink signal $\textbf{x}$ to two types of users equipped with a single antenna.
Here, the first type of user is the delay-tolerant user and the second type of user is the delay-constrained user.
The delay-tolerant user means a user who wants to achieve high spectral efficiency without the target latency.
On the contrary, the delay-constrained user means a user who needs to receive the downlink signal within the target latency.
The numbers of delay-tolerant users and delay-constrained users are denoted as $K_t$ and $K_s$, respectively.
For convenience, we then define the total user set as $\mathcal{K}=\{1,2,\cdots,K_t+K_s\}$, which is divided into the delay-tolerant user set $\mathcal{K}_t=\{1,2,\cdots,K_t\}$ and the delay-constrained user set $\mathcal{K}_s=\{K_t+1,K_t+2,\cdots,K_t+K_s\}$ (i.e., $\mathcal{K} = \mathcal{K}_t \cup \mathcal{K}_s$).
%
%
%
%\/\/\/\/\/\/\/\/\/\/\/\/\/\/\/\/\/\/\/\/\/\/\/\/\/\/\/\/\/\/\/\/\/\/\/\/\/\/\/\/\/\/\/\/\/\/\/\/\/\/\/\/\/\/\/\/\/\/\/\/\/\/\/\/\/\/\/\/\/\/\/\/\/\/\/\/\/\/\/\/\/\/\/\/\/\/\/
%***** x-axis:  [tc][bc][0.7] y-axis: [bc][tc][0.7], legend: [Bl][Bl][0.59]
%\/\/\/\/\/\/\/\/\/\/\/\/\/\/\/\/\/\/\/\/\/\/\/\/\/\/\/\/\/\/\/\/\/\/\/\/\/\/\/\/\/\/\/\/\/\/\/\/\/\/\/\/\/\/\/\/\/\/\/\/\/\/\/\/\/\/\/\/\/\/\/\/\/\/\/\/\/\/\/\/\/\/\/\/\/\/\/
%
%
%
In this network, we express the downlink signal $\textbf{x}$ as
\begin{align}
\textbf{x} = \sum_{k=1}^{K_t+K_s} \textbf{u}_k s_k \label{eq:downlink_signl}
\end{align}
where $\textbf{u}_k$ is the precoding vector for the user $k$ and $s_k$ is the transmit symbol for the user $k$ with the average power $P=\mathbb{E}\left[\left|s_k\right|^2\right]$.
%In \eqref{eq:downlink_signl}, the user indices of the delay-tolerant users are $k = 1, \cdots, K_t \in \mathcal{K}_t$ and those of the delay-constrained users are $k = K_t+1, \cdots, K_t+K_s \in \mathcal{K}_s$.
We assume that the \ac{BS} determines the precoding vector for one resource frame and knows the perfect \ac{CSI} of all the users.
From \eqref{eq:downlink_signl}, the received signal at the user $k$ is given by
\begin{align}
y_k = \textbf{h}_k^\text{H} \textbf{u}_k s_k + \sum_{i \neq k}^{K_t+K_s} \textbf{h}_k^\text{H} \textbf{u}_i s_i  + z_k , \quad \forall k \label{eq:signal}
\end{align}
where $z_k$ is the noise signal at the user $k$, distributed as $\mathcal{CN}(0,\sigma_k^2)$.
In \eqref{eq:signal}, $\textbf{h}_k = \left[h_k^1, h_k^2, \cdots, h_k^N \right] \in \mathbb{C}^{N \times 1}$ is the channel vector from the \ac{BS} to the user $k$.
The distribution of $\textbf{h}_k$ is considered as the complex Gaussian, i.e., $\textbf{h}_k \sim \mathcal{CN}(0,\textbf{C}_k)$ where $\textbf{C}_k=\mathbb{E} \left[\textbf{h}_k \textbf{h}_k^\text{H}\right] \in \mathbb{C}^{N \times N}$ is the spatial covariance matrix of the channel.

\subsection{Network Model with IR-HARQ scheme}
In the DCTU-MIMO network, to increase the reliability of the delay-constrained users, the \ac{HARQ} is used, which is a combination of \ac{FEC} and \ac{ARQ} schemes, among which we adopt the \ac{IR-HARQ} scheme \cite{Cai:01}.
The \ac{IR-HARQ} scheme is one of the \ac{HARQ} schemes that the erroneous decoded packets are combined with incremental redundancy after each retransmission.
Specifically, when a \ac{BS} transmits the downlink signal to a user $k$, the user $k$ sends an \ac{ACK} or a \ac{NACK} message to the \ac{BS}.
If the \ac{BS} receives a \ac{ACK} message, the \ac{BS} transmits new downlink signal to the user $k$.
On the other hand, if the \ac{BS} receives a \ac{NACK} message, the \ac{BS} retransmits the same downlink signal with additional redundancy bits and the user $k$ combines the signal, received during retransmissions for the same downlink signal, and uses it for decoding.
In this consideration, we assume a resource frame is divided into $L$ time slots.
The user scheduling for delay-tolerant users is done for $L$ time slots, while that for delay-constrained users is done for $T$ time slots ($T<L$) to meet the \ac{URLLC} latency requirement.
%We also assume that the downlink signals from $T$ transmission rounds are jointly decoded at the user $k$.

Under the same circumstances as the DCTU-MIMO network considered in Section~\ref{subsec:system_without}, given the \ac{IR-HARQ} scheme, the downlink signal at the $t$-th transmission round is expressed as
\begin{align}
	\textbf{x}(t) = \sum_{k=1}^{K_t+K_s} \textbf{u}_k(t) s_k , \quad \forall k , \, t=1,\cdots,T \label{eq:downlink_signl_IR}
\end{align}
where $\textbf{u}_k(t)\in \mathbb{C}^{N \times 1}$ is the precoding vector of the user $k$ at the $t$-th transmission round.
From \eqref{eq:downlink_signl_IR}, the received signal of the user $k$ at the $t$-th transmission round is given by
\begin{align}
	y_k(t) = \textbf{h}_k^\text{H}(t) \textbf{u}_k(t) s_k + \sum_{i \neq k}^{K_t+K_s} \textbf{h}_k^\text{H}(t) \textbf{u}_i(t) s_i  + z_k  , \quad \forall k , \, t=1,\cdots,T \label{eq:signal_IR}
\end{align}
where $\textbf{h}_k(t) \in \mathbb{C}^{N \times 1}$ is the channel vector from the \ac{BS} to the user $k$ at the $t$-th transmission round.
The distribution of $\textbf{h}_k(t)$ is also considered as the complex Gaussian, i.e., $\textbf{h}_k(t) \sim \mathcal{CN}(0,\textbf{C}_k(t))$ where $\textbf{C}_k(t)=\mathbb{E} \left[\textbf{h}_k(t) \textbf{h}_k^\text{H}(t)\right] \in \mathbb{C}^{N \times N}$ is the spatial covariance matrix of the channel at the $t$-th transmission round.

\section{Spectral Efficiency Analysis} \label{sec:SE}
In this section, we analyze the spectral efficiency without and with the \ac{IR-HARQ} scheme.
Firstly, the delay-tolerant users can be allocated to long enough blocklength because the \ac{BS} allocates more time resources to the delay-tolerant users. Therefore, for the delay-tolerant users, we consider the spectral efficiency using the infinite blocklength coding, which follows the Shannon capacity.
For delay-constrained users, unlike delay-tolerant users, we consider the spectral efficiency in the finite blocklength regime.
This is because, to satisfy the \ac{URLLC} latency requirement, the delay-constrained users are allocated small time resources.

\subsection{Spectral Efficiency without \ac{IR-HARQ} scheme} \label{subsec:SE_A}
From \eqref{eq:signal}, the spectral efficiency of the delay-tolerant user, which follows the Shannon capacity, can be expressed as
\begin{align}
R_k(\gamma_k) = \log_2(1 + \gamma_k) , \quad  k \in \mathcal{K}_t \label{eq:rate_t}
\end{align}
where $\gamma_k$ is the \ac{SINR} of the user $k$, given by
\begin{align}
\gamma_k = 
\frac{\left| \textbf{h}_k^\text{H} \textbf{u}_k \right|^2} 
{\sum_{i \neq k}^{K_t+K_s} \left|\textbf{h}_k^\text{H} \textbf{u}_i \right|^2 + \frac{\sigma_k^2}{P}}
\end{align}
for the noise signal power $\sigma_k^2$.
To make the spectral efficiency as the Rayleigh quotient that is a suitable form for applying the \ac{GPI} method, we define the network-wide precoding vector $\textbf{u}$ as
\begin{align}
\textbf{u} = \left[\underbrace{\textbf{u}_1^\text{H},\cdots,\textbf{u}_{K_t}^\text{H}}_{\text{delay-tolerant users}},\underbrace{\textbf{u}_{K_t+1}^\text{H},\cdots,\textbf{u}_{K_t+K_s}^\text{H}}_{\text{delay-constrained users}}\right]^\text{H} \in \mathbb{C}^{N(K_t+K_s) \times 1} . \label{eq:precoding_net}
\end{align}
Since the spectral efficiency is an increasing function with the transmit power consumption $\left\lVert \textbf{u} \right\rVert$, we assume that the maximum transmit power is used, i.e., $\left\lVert \textbf{u} \right\rVert=1$.
Using \eqref{eq:precoding_net}, we rewrite the spectral efficiency of the delay-tolerant user as
\begin{align}
R_k(\textbf{u})
&= \log_2 \left(\frac{\sum_{i=1}^{K_t+K_s} \textbf{u}_i^\text{H} \left(\textbf{h}_k \textbf{h}_k^\text{H}\right) \textbf{u}_i + \frac{\sigma_k^2}{P}} 
{\sum_{i \neq k}^{K_t+K_s}\textbf{u}_i^\text{H} \left(\textbf{h}_k \textbf{h}_k^\text{H}\right) \textbf{u}_i + \frac{\sigma_k^2}{P}} \right) \nonumber \\
&= \log_2  \left(\frac{\textbf{u}^\text{H} \textbf{A}_k \textbf{u}} {\textbf{u}^\text{H} \textbf{B}_k \textbf{u}}\right) , \quad  k \in \mathcal{K}_t
\label{eq:rate_net}
\end{align}
where $\textbf{A}_k = \text{diag} \left(\textbf{h}_k \textbf{h}_k^\text{H}, \cdots , \textbf{h}_k \textbf{h}_k^\text{H}\right) + \frac{\sigma_k^2}{P} \textbf{I}_{N(K_t+K_s)}$ and $\textbf{B}_k = \textbf{A}_k - \text{diag} \left(\textbf{0},\cdots,\textbf{h}_k \textbf{h}_k^\text{H},\cdots,\textbf{0}\right)$ is constructed by subtracting the $k$-th sub-block matrix from $\textbf{A}_k$.

For delay-constrained users, for given blocklength $m$ and target decoding error probability $\varepsilon_k$, the spectral efficiency is expressed as \cite{Poy:10,Yan:14}
\begin{align}
R_k(\gamma_k) = \log_2(1 + \gamma_k) - \sqrt{\frac{V(\gamma_k)} {m}} Q^{-1}(\varepsilon_k) , \quad  k \in \mathcal{K}_s \label{eq:rate_s}
\end{align}
where $V(\gamma_k)$ is the channel dispersion of the user $k$, which are given by
\begin{align}
V(\gamma_k) = 
\frac{2 \gamma_k} {1 + \gamma_k} \left(\log_2 e\right)^2 , \quad  k \in \mathcal{K}_s .
\end{align}
Note that as the inter-user interference is considered, the channel dispersion $V(\gamma_k)$ cannot be defined in \ac{AWGN} channel.
Hence, we use the channel dispersion of Gaussian codebooks under the non-Gaussian noise and the nearest-neighbor decoding \cite{Sca:17}.
However, since the spectral efficiency of delay-constrained users consists of the mixture of the log function and the square root function, it is difficult to deal with an optimization problem that maximizes the sum spectral efficiency of delay-tolerant users under latency constraints.
Hence, to make the spectral efficiency of delay-constrained users a tractable form, we present the upper bound of the channel dispersion using the similar approach in \cite{Ren:20} in the following Lemma.
\begin{lemma} \label{lemma:bound}
For any given $\tilde{x} > 0$, the upper bound of $\sqrt{\frac{2 x} {1 + x}}$ can be given as
\begin{align}
\sqrt{\frac{2 x} {1 + x}} \le
\rho(\tilde{x}) \ln (1+x) + \eta(\tilde{x}) , \quad \forall x > 0 \label{eq:dispersion}
\end{align}
where $\rho(\tilde{x}) \hspace{-0.8mm} = \hspace{-0.8mm} \frac{1} {\sqrt{2 \tilde{x} \left(1+\tilde{x}\right)}}$ and $\eta(\tilde{x}) \hspace{-0.8mm} = \hspace{-0.8mm} \sqrt{\frac{2 \tilde{x}} {1 + \tilde{x}}} \hspace{-0.5mm} - \hspace{-0.5mm} \rho(\tilde{x}) \ln (1 \hspace{-0.4mm} + \hspace{-0.4mm} \tilde{x})$.
When $x\hspace{-0.8mm}=\hspace{-0.8mm}\tilde{x}$, the equality in \eqref{eq:dispersion} holds.
\end{lemma}
\begin{IEEEproof}
	Define $H(x) = \sqrt{\frac{2 x} {1 + x}}$ and $G(x,\tilde{x}) = \rho(\tilde{x}) \ln (1+x) + \eta(\tilde{x})$.
	Firstly, we can readily know that $H(x)$ and $G(x,\tilde{x})$ are the same at $x=\tilde{x}$ by substituting given $\rho(\tilde{x})$ and $\eta(\tilde{x})$ into $G(x,\tilde{x})$, which proves the equality of \eqref{eq:dispersion}.
	We then prove the inequality $H(x) < G(x,\tilde{x})$ by defining a function $F(x,\tilde{x}) \triangleq G(x,\tilde{x}) - H(x)$ and obtaining the first derivative of $F(x,\tilde{x})$ with respect to $x$.
	The first derivative of $F(x,\tilde{x})$ with respect to $x$ can be given by
	\begin{align}
	\frac{\partial F(x,\tilde{x})} {\partial x} 
	&= \frac{(1+\tilde{x})\sqrt{2x}\sqrt{(1+x)^3} - (1+x)\sqrt{2\tilde{x}}\sqrt{(1+\tilde{x})^3}} {(1+x)\sqrt{2x}\sqrt{(1+x)^3}\sqrt{2\tilde{x}}\sqrt{(1+\tilde{x})^3}}. \label{eq:dev}
	\end{align}
	Since $x$ and $\tilde{x}$ are positive values, the sign of $\frac{\partial F(x,\tilde{x})} {\partial x}$ is affected by the numerator in \eqref{eq:dev}.
	The numerator can be represented by
	\begin{align}
	J(x,\tilde{x}) 
	= (1+\tilde{x})(1+x) \left(
		\sqrt{2x}\sqrt{(1+x)} - \sqrt{2\tilde{x}}\sqrt{(1+\tilde{x})}
		\right). \label{eq:numerator}
	\end{align}
	From new, we will show that $J(x,\tilde{x}) < 0$ when $0 < x < \tilde{x}$ and $J(x,\tilde{x}) > 0$ when $x > \tilde{x}$.
	First, let us define $U(x) \triangleq \sqrt{2x}\sqrt{1+x}$ and obtain the first derivative of $U(x)$ with respect to $x$ as
	\begin{align}
	\frac{\partial U(x)} {\partial x} 
	&= \frac{1+2x}{\sqrt{2x(1+x)}}. \label{eq:dev_num}
	\end{align}
	From \eqref{eq:dev_num}, we can know that $\frac{\partial U(x)} {\partial x} > 0$ when $x>0$.
	Hence, when $0 < x < \tilde{x}$, we have $J(x,\tilde{x}) < 0$ because $U(x) < U(\tilde{x})$. Consequently, we have $\frac{\partial F(x,\tilde{x})} {\partial x} < 0$, which means $F(x,\tilde{x})$ is a monotonically decreasing function of $x$.
	Hence, the inequality $F(x,\tilde{x}) < F(0,\tilde{x}) = \eta(\tilde{x})$ can be satisfied.
	Here, since $\eta(\tilde{x}) > 0 $ for $\tilde{x} > 0$ and $F(\tilde{x}, \tilde{x})=0$, we obtain $F(x,\tilde{x}) > 0 $ holds, which results in $G(x,\tilde{x}) > H(x)$ when $0 < x < \tilde{x}$.
	
	On the other hand, for $x > \tilde{x}$, we have $U(x) > U(\tilde{x})$ and $J(x,\tilde{x}) > 0$.
	Therefore, $\frac{\partial F(x,\tilde{x})} {\partial x} > 0$, which signifies $F\hspace{-0.3mm}(\hspace{-0.2mm}x\hspace{-0.2mm},\hspace{-0.3mm}\tilde{x}\hspace{-0.2mm})$ is a monotonically increasing function of $x$.
	Hence, we have $F\hspace{-0.3mm}(\hspace{-0.2mm}x\hspace{-0.2mm},\hspace{-0.3mm}\tilde{x}\hspace{-0.2mm}) \hspace{-1.2mm} > \hspace{-1mm} \eta(\tilde{x})$ and $G\hspace{-0.3mm}(\hspace{-0.2mm}x\hspace{-0.2mm},\hspace{-0.3mm}\tilde{x}\hspace{-0.2mm}) \hspace{-1.2mm} > \hspace{-1mm} H\hspace{-0.3mm}(x)$ when $x \hspace{-1mm} > \hspace{-0.8mm} \tilde{x}$.
	Therefore, when $x\hspace{-1mm}>\hspace{-0.8mm}0$, $G\hspace{-0.3mm}(\hspace{-0.2mm}x\hspace{-0.2mm},\hspace{-0.3mm}\tilde{x}\hspace{-0.2mm})$ is always greater than $H\hspace{-0.3mm}(x)$.
\end{IEEEproof}

From Lemma \ref{lemma:bound}, 
by using \eqref{eq:dispersion} into \eqref{eq:rate_s}, for given $\tilde{\gamma}_k$, the lower bound of the spectral efficiency for the delay-constrained user can be represented as
\begin{align}
R_k(\gamma_k) \ge \tilde{R}_k(\gamma_k, \tilde{\gamma}_k) &=
\log_2(1 + \gamma_k) - \frac{Q^{-1}(\varepsilon_k)} {\sqrt{m}}  \left\{\rho(\tilde{\gamma}_k) \log_2 (1+\gamma_k) + \eta(\tilde{\gamma}_k) \log_2 e\right\}  \nonumber \\
%
%&= \log_2 \left\{ 
%\frac{1 + \gamma_k} {(1+\gamma_k)^{f_k\left(\tilde{\gamma}_k\right)} } 
%\right\} - g_k\left(\tilde{\gamma}_k\right)  \nonumber \\
%
&= \log_2  \left\{ 
\left(\frac{\sum_{i=1}^{K_t+K_s} \left|\textbf{h}_k^\text{H} \textbf{u}_i \right|^2  +  \frac{\sigma_k^2}{P}} 
{\sum_{i \neq k}^{K_t+K_s} \left|\textbf{h}_k^\text{H} \textbf{u}_i \right|^2 + \frac{\sigma_k^2}{P}}\right)^{1-f_k\left(\tilde{\gamma}_k\right)} 
\right\} 
-  g_k\left(\tilde{\gamma}_k\right) , \quad  k \in \mathcal{K}_s
\label{eq:rate_s_eq}
\end{align}
where $\rho_k(\tilde{\gamma}_k) = \frac{1} {\sqrt{2 \tilde{\gamma}_k \left(1+\tilde{\gamma}_k\right)}}$, $\eta_k(\tilde{\gamma}_k) = \sqrt{\frac{2 \tilde{\gamma}_k} {1 + \tilde{\gamma}_k}} - \rho_k(\tilde{\gamma}_k) \ln \left(1 + \tilde{\gamma}_k\right)$, $f_k\left(\tilde{\gamma}_k\right)=\frac{Q^{-1}(\varepsilon_k) \rho_k(\tilde{\gamma}_k)}{\sqrt{m}}$, and $g_k\left(\tilde{\gamma}_k\right)=\frac{Q^{-1}(\varepsilon_k)\eta_k(\tilde{\gamma}_k)} {\sqrt{m}} \log_2e$.
Further, using \eqref{eq:precoding_net}, we rewrite the lower bound of the spectral efficiency for the delay-constrained user as a function of $\textbf{u}$ and $\tilde{\gamma}_k$ as
\begin{align}
\tilde{R}_k(\textbf{u}, \tilde{\gamma}_k)
= \log_2 \left\{  \left(\frac{\textbf{u}^\text{H} \textbf{A}_k \textbf{u}} {\textbf{u}^\text{H} \textbf{B}_k \textbf{u}}\right)^{1 - f_k\left(\tilde{\gamma}_k\right)} \right\} - g_k\left(\tilde{\gamma}_k\right) , \quad  k \in \mathcal{K}_s .
\label{eq:rate_s_net}
\end{align}

Using \eqref{eq:rate_s_net}, we can also present the upper bound of the communication latency, given by
\begin{align}
\frac{D_s}{\tilde{R}_k(\textbf{u}, \tilde{\gamma}_k)} =  \frac{D_s} {\log_2 \left\{  \left(\frac{\textbf{u}^\text{H} \textbf{A}_k \textbf{u}} {\textbf{u}^\text{H} \textbf{B}_k \textbf{u}}\right)^{1 - f_k\left(\tilde{\gamma}_k\right)} \right\} - g_k\left(\tilde{\gamma}_k\right)} , \quad  k \in \mathcal{K}_s \label{eq:latency}
\end{align}
where $D_s$ is the data size of the downlink signal.
%Note that it is reasonable to analyze the the upper bound of the communication latency because if the upper bound of the communication latency is satisfied, the actual communication latency is also satisfied.
%In addition, when we select proper large value of $\tilde{\gamma}_k$, the difference between the actual communication latency and the upper bound of the communication latency is small.

\subsection{Spectral Efficiency with \ac{IR-HARQ} scheme} \label{subsec:SE_B}
Let us define $\textbf{u}_\text{IR}$ as the network-wide precoding vector of the \ac{IR-HARQ} scheme, which can be expressed as $\textbf{u}_\text{IR} = \left[\underbrace{\textbf{u}_1^\text{H}(1),\cdots,\textbf{u}_{K_t+K_s}^\text{H}(1)}_{\textbf{u}_\text{IR}^\text{H}(1)},\cdots,\underbrace{\textbf{u}_1^\text{H}(T),\cdots,\textbf{u}_{K_t+K_s}^\text{H}(T)}_{\textbf{u}_\text{IR}^\text{H}(T)}\right]^\text{H} \in \mathbb{C}^{N(K_t+K_s)T \times 1}$,
where $\textbf{u}_\text{IR}(t)=[\textbf{u}_1^\text{H}(t),\cdots,\textbf{u}_{K_t+K_s}^\text{H}(t)]^\text{H}\in \mathbb{C}^{N(K_t+K_s) \times 1}$ is the network-wide precoding vector of the \ac{IR-HARQ} scheme at the $t$-th transmission round.
From \eqref{eq:signal_IR}, for given $T$ transmission rounds, since we use the maximum transmit power, i.e., $\left\lVert \textbf{u}_\text{IR} \right\rVert = 1$, spectral efficiencies of the delay-tolerant user and the delay-constrained user are expressed as \cite{Cai:01,Pol:09}
\begin{align}
	R_k^\text{IR}(\textbf{u}_\text{IR})
	&= \sum_{t=1}^{T} R_{k,t}(\textbf{u}_\text{IR})
	= \sum_{t=1}^{T} \log_2 \left(1 + \gamma_k(t)\right) 
	= \sum_{t=1}^{T} \log_2 \left(\frac{\textbf{u}_\text{IR}^\text{H} \textbf{D}_k(t) \textbf{u}_\text{IR}} {\textbf{u}_\text{IR}^\text{H} \textbf{E}_k(t) \textbf{u}_\text{IR}}\right) , \quad  k \in \mathcal{K}_t \, , \nonumber \\
	R_k^\text{IR}(\textbf{u}_\text{IR})
	&= \sum_{t=1}^{T} \log_2 \left(1 + \gamma_k(t)\right) 
	- \sqrt{\frac{\sum_{t=1}^{T} V(\gamma_k(t))} {m}} Q^{-1}(\varepsilon_k) \nonumber \\
	&= \sum_{t=1}^{T} \log_2 \left(\frac{\textbf{u}_\text{IR}^\text{H} \textbf{D}_k(t) \textbf{u}_\text{IR}} {\textbf{u}_\text{IR}^\text{H} \textbf{E}_k(t) \textbf{u}_\text{IR}}\right) 
	- \sqrt{\frac{\sum_{t=1}^{T} V(\gamma_k(t))} {m}} Q^{-1}(\varepsilon_k) , \quad  k \in \mathcal{K}_s \label{eq:rate_HARQ}
\end{align}
where $R_{k,t}(\textbf{u}_\text{IR})$ is the spectral efficiency of user $k$ at the $t$-th transmission round and $\textbf{D}_k(t)$ is given by
\begin{align}
	\textbf{D}_k(t) = \begin{bmatrix}
			\textbf{0}       & \dots    &   \textbf{0}       & \dots     &   \textbf{0}       \\
			\vdots  &  \ddots   &   \vdots  &      &   \vdots \\
			\textbf{0}       & \dots    &   \textbf{D}_k^t(t)       & \dots     &   \textbf{0}       \\
			\vdots  &     &    \vdots   &  \ddots  &   \vdots  \\   
			\textbf{0}    & \dots     &   \textbf{0}       & \dots     &   \textbf{0}       \\
		\end{bmatrix}
	  + \frac{\sigma_k^2}{P} \textbf{I}_{N(K_t+K_s)T} \in \mathbb{C}^{N(K_t+K_s)T \times N(K_t+K_s)T} \label{eq:matrix_D}
\end{align}
where $\textbf{D}_k^t(t) \hspace{-0.7mm} = \hspace{-0.7mm} \text{diag} \left(\textbf{h}_k(t) \textbf{h}_k^\text{H}(t) , \cdots, \textbf{h}_k(t) \textbf{h}_k^\text{H}(t)\right) \hspace{-0.7mm} \in \hspace{-0.7mm} \mathbb{C}^{N(K_t+K_s) \times N(K_t+K_s)}$ is $t$-th sub-block matrix of $\textbf{D}_k(t)$.
In \eqref{eq:rate_HARQ}, $\textbf{E}_k(t) = \textbf{D}_k(t) - \text{diag} \left(\textbf{0},\cdots,\textbf{h}_k(t) \textbf{h}_k^\text{H}(t),\cdots,\textbf{0}\right) \in \mathbb{C}^{N(K_t+K_s)T \times N(K_t+K_s)T}$ is constructed by subtracting the $k$-th sub-block matrix from $\textbf{D}_k^t(t)$ and $\gamma_k(t)$ is the \ac{SINR} of user $k$ at the $t$-th transmission round, given by
\begin{align}
	\gamma_k(t) 
	= \frac{\left| \textbf{h}_k^\text{H}(t) \textbf{u}_k(t) \right|^2} 
	{\sum_{i \neq k}^{K_t+K_s} \left|\textbf{h}_k^\text{H}(t) \textbf{u}_i(t) \right|^2 + \frac{\sigma_k^2}{P}}  , \quad \forall k , \, \forall t  . \label{eq:SINR_IR}
\end{align}

As mentioned in Section~\ref{subsec:SE_A}, since the spectral efficiency of delay-constrained users consists of the mixture of the log function and the square root function, it is difficult to deal with the optimization problem that maximizes the sum spectral efficiency of delay-tolerant users under the latency constraint.
Hence, to make the spectral efficiency of delay-constrained users as a tractable form, for any given $\tilde{\gamma}_k(t) > 0$, we provide the upper bound of $\sqrt{\sum_{t=1}^{T} \frac{2 \gamma_k(t)} {1 + \gamma_k(t)}}$ as
\begin{align}
	\sqrt{\sum_{t=1}^{T} 
		\frac{2 \gamma_k(t)} {1 + \gamma_k(t)}} 
	\overset{\underset{\mathrm{(a)}}{}}{\le} \sum_{t=1}^{T} 
	\sqrt{\frac{2 \gamma_k(t)} {1 + \gamma_k(t)}} 
	\overset{\underset{\mathrm{(b)}}{}}{\le} \sum_{t=1}^{T} 
	\left\{
	\rho(\tilde{\gamma}_k(t)) \ln (1 + \gamma_k(t)) + \eta(\tilde{\gamma}_k(t))
	\right\} \label{eq:inequality_HARQ}
\end{align}
where (a) is due to the triangle inequality, (b) is obtained using \eqref{eq:dispersion}, $\rho(\tilde{\gamma}_k(t)) = \frac{1} {\sqrt{2 \tilde{\gamma}_k(t) \left(1 + \tilde{\gamma}_k(t)\right)}}$, and $\eta(\tilde{\gamma}_k(t)) = \sqrt{\frac{2 \tilde{\gamma}_k(t)} {1 + \tilde{\gamma}_k(t)}} - \rho(\tilde{\gamma}_k(t)) \ln (1 + \tilde{\gamma}_k(t))$.
By using \eqref{eq:inequality_HARQ} into \eqref{eq:rate_HARQ}, we obtain the lower bound of the spectral efficiency for delay-constrained users as
\begin{align}
	R_k^\text{IR}(\textbf{u}_\text{IR}) \ge
	\tilde{R}_k^\text{IR}\hspace{-0.2mm}(\textbf{u}_\text{IR}, \hspace{-0.2mm}\tilde{\gamma}_k(1), \cdots, \hspace{-0.2mm} \tilde{\gamma}_k(T)) 
	&= 
	\sum_{t=1}^{T} \hspace{-0.5mm} \left[ 
	\log_2 \hspace{-0.7mm} \left\{ \hspace{-1.2mm}
	\left(\hspace{-0.3mm}\frac{\textbf{u}_\text{IR}^\text{H} \textbf{D}_k\hspace{-0.2mm}(t) \textbf{u}_\text{IR}} {\textbf{u}_\text{IR}^\text{H} \textbf{E}_k\hspace{-0.2mm}(t) \textbf{u}_\text{IR}}\hspace{-0.7mm}\right)^{\hspace{-1mm}1-f_{k,t}\left(\tilde{\gamma}_k(t)\right)} \hspace{-0.5mm} \right\}  
	\hspace{-0.5mm} - \hspace{-0.3mm} g_{k,t}\left(\tilde{\gamma}_k(t)\right)
	\right] \nonumber \\
	&= 
	 \sum_{t=1}^{T} \tilde{R}_{k,t}(\textbf{u}_\text{IR}, \tilde{\gamma}_k(t)), \quad k \in \mathcal{K}_s \hspace{-1mm} \label{eq:rate_s_HARQ_bound}
\end{align}
where $\tilde{R}_{k,t}(\textbf{u}_\text{IR}, \tilde{\gamma}_k(t))$ is the lower bound of spectral efficiency of user $k$ at the $t$-th transmission round, $f_{k,t}\left(\tilde{\gamma}_k(t)\right)=\frac{Q^{-1}(\varepsilon_k) \rho_k(\tilde{\gamma}_k(t))}{\sqrt{m}}$, and $g_{k,t}\left(\tilde{\gamma}_k(t)\right)=\frac{Q^{-1}(\varepsilon_k)\eta_k(\tilde{\gamma}_k(t))} {\sqrt{m}} \log_2e$.

However, since the \ac{BS} does not know the future \ac{CSI} of all the users in practice, we cannot know the exact spectral efficiency of future transmission rounds.
Therefore, future transmission rounds, by estimating the channel covariance matrix from geometrical locations or the \ac{AoA} of users, we obtain the approximation of the ergodic spectral efficiencies as
	\begin{align}
		&\mathbb{E}_{\textbf{h}_k(t)}\left[R_{k,t}(\textbf{u}_\text{IR})\right] = 
		\mathbb{E}_{\textbf{h}_k(t)}\left[\log_2 \left(1 + \frac{\left| \textbf{h}_k^\text{H}(t) \textbf{u}_k(t) \right|^2} 
		{\sum_{i \neq k}^{K_t+K_s} \left|\textbf{h}_k^\text{H}(t) \textbf{u}_i(t) \right|^2 + \frac{\sigma_k^2}{P}}\right)\right] \nonumber \\
		&\overset{\underset{\mathrm{(a)}}{}}{\approx} \hspace{-0.5mm}
		\log_2 \hspace{-1mm} \left(\hspace{-0.5mm} 1 \hspace{-0.5mm} + \hspace{-0.5mm} \frac{\mathbb{E}_{\textbf{h}_k(t)}\left[\left| \textbf{h}_k^\text{H}(t) \textbf{u}_k(t) \right|^2\right]} 
		{\mathbb{E}_{\textbf{h}_k(t)} \hspace{-0.5mm} \left[\sum_{i \neq k}^{K_t+K_s} \hspace{-0.3mm} \left|\textbf{h}_k^\text{H}(t) \textbf{u}_i(t) \right|^2\right] \hspace{-0.5mm} + \hspace{-0.5mm} \frac{\sigma_k^2}{P}}\right) \hspace{-0.7mm}
		= \hspace{-0.5mm} \log_2 \hspace{-0.5mm} \left(\frac{\textbf{u}_\text{IR}^\text{H} \tilde{\textbf{D}}_k(t) \textbf{u}_\text{IR}} {\textbf{u}_\text{IR}^\text{H} \tilde{\textbf{E}}_k(t) \textbf{u}_\text{IR}} \hspace{-0.5mm} \right) \hspace{-0.5mm}
		= \hspace{-0.5mm} \hat{R}_{k,t}(\textbf{u}_\text{IR}) ,  \,\,  k \in \mathcal{K}_t , \nonumber \\
		&\mathbb{E}_{\textbf{h}_k\hspace{-0.3mm}(t)}\hspace{-1.3mm}\left[\hspace{-0.3mm}\tilde{R}_{k,t}\hspace{-0.3mm}(\textbf{u}_\text{IR}\hspace{-0.2mm},\hspace{-0.4mm} \tilde{\gamma}_k(\hspace{-0.3mm}t\hspace{-0.3mm})\hspace{-0.5mm})\right] \hspace{-0.7mm}
		%&= \mathbb{E}_{\textbf{h}_k(t)}\hspace{-1.5mm}\left[\log_2 \hspace{-1mm}\left\{ \hspace{-1.5mm}
		%\left(\hspace{-0.5mm}1 \hspace{-0.9mm} + \hspace{-0.8mm} \frac{\left| \textbf{h}_k^\text{H}(t) \textbf{u}_k(t) \right|^2} 
		%{\sum_{i \neq k}^{K_t+K_s} \hspace{-0.5mm} \left|\textbf{h}_k^\text{H}(t) \textbf{u}_i(t) \right|^2 \hspace{-0.9mm} + \hspace{-0.7mm} \frac{\sigma_k^2}{P}}\hspace{-0.5mm}\right)^{\hspace{-1.2mm}1-f_{k,t}\left(\tilde{\gamma}_k(t)\right)}
		%\right\} \hspace{-0.8mm} - \hspace{-0.7mm}  g_{k,t}\hspace{-0.8mm}\left(\tilde{\gamma}_k(t)\right)\hspace{-0.5mm}\right] \nonumber \\
		%
		%&\overset{\underset{\mathrm{(a)}}{}}{\approx} \log_2 \hspace{-1mm}\left\{ \hspace{-1.8mm}
		%\left(\hspace{-0.7mm}1 \hspace{-0.9mm} + \hspace{-0.8mm} \frac{\mathbb{E}_{\textbf{h}_k(t)}\hspace{-1.5mm}\left[\left| \textbf{h}_k^\text{H}(t) \textbf{u}_k(t) \right|^2\right]}
		%{\mathbb{E}_{\textbf{h}_k(t)}\hspace{-1.5mm}\left[\sum_{i \neq k}^{K_t+K_s} \hspace{-0.5mm} \left|\textbf{h}_k^\text{H}(t) \textbf{u}_i(t) \right|^2\right] \hspace{-1mm} + \hspace{-0.7mm} \frac{\sigma_k^2}{P}}\hspace{-0.5mm}\right)^{\hspace{-1.2mm}1-f_{k,t}\left(\tilde{\gamma}_k(t)\right)}
		%\right\} \hspace{-0.8mm} - \hspace{-0.7mm}  g_{k,t}\hspace{-0.8mm}\left(\tilde{\gamma}_k(t)\right) \nonumber \\
		%
		\overset{\underset{\mathrm{(a)}}{}}{\approx} \log_2 \hspace{-1mm} \left\{ \hspace{-2mm}
		\left(\hspace{-0.8mm}\frac{\textbf{u}_\text{IR}^\text{H} \tilde{\textbf{D}}_k\hspace{-0.4mm}(\hspace{-0.3mm}t\hspace{-0.4mm}) \textbf{u}_\text{IR}} {\textbf{u}_\text{IR}^\text{H} \tilde{\textbf{E}}_k\hspace{-0.3mm}(\hspace{-0.3mm}t\hspace{-0.4mm}) \textbf{u}_\text{IR}}\hspace{-0.7mm}\right)^{\hspace{-1.2mm}1\hspace{-0.2mm}-\hspace{-0.3mm}f_{k,t}\left(\tilde{\gamma}_k(\hspace{-0.2mm}t\hspace{-0.1mm})\hspace{-0.2mm}\right)} \hspace{-0.7mm}
		\right\} \hspace{-1.2mm} - \hspace{-0.8mm}  g_{k,t}\hspace{-1mm}\left(\hspace{-0.2mm}\tilde{\gamma}_k\hspace{-0.3mm}(\hspace{-0.3mm}t\hspace{-0.3mm})\hspace{-0.4mm}\right) \hspace{-0.8mm} = \hspace{-0.8mm} \hat{R}_{k,t}\hspace{-0.4mm}(\textbf{u}_\text{IR}\hspace{-0.2mm}, \hspace{-0.4mm} \tilde{\gamma}_k\hspace{-0.3mm}(\hspace{-0.3mm}t\hspace{-0.3mm})\hspace{-0.4mm}) , \,\, k \hspace{-0.6mm} \in \hspace{-0.6mm} \mathcal{K}_s \hspace{-2mm} \label{eq:rate_approx}
	\end{align}
	where (a) is from the Lemma 1 of \cite{Zha:14}. 
	In \eqref{eq:rate_approx}, $\tilde{\textbf{D}}_k(t)$ and $\tilde{\textbf{E}}_k(t)$ are obtained by replacing $\textbf{h}_k(t) \textbf{h}_k^\text{H}(t)$ of $\textbf{D}_k(t)$ and $\textbf{E}_k(t)$ with $\textbf{C}_k(t)$.
	Here, $\textbf{C}_k(t)=\mathbb{E} \left[\textbf{h}_k(t) \textbf{h}_k^\text{H}(t)\right] \in \mathbb{C}^{N \times N}$ is the spatial covariance matrix of the channel of user $k$ at the $t$-th transmission round.

%
%%---------------------------------------------------------------------------%
%%                        Sec: Precoding Algorithm                           %
%%---------------------------------------------------------------------------%
%
\section{Precoding Design for DCTU-MIMO network without \ac{IR-HARQ} scheme} \label{sec:Algorithm}
In this section, we first formulate an optimization problem that maximizes the spectral efficiency of delay-tolerant users while satisfying the communication latency constraint of delay-constrained users for a DCTU-MIMO network.
We then derive the first-order optimality condition of the optimization problem and propose the computationally efficient algorithm to find a sub-optimal solution that satisfies the first-order optimality condition.

\subsection{Problem Formulation}
Under the communication latency constraint, the sum spectral efficiency maximization problem can be formulated as
\begin{align}
	&\underset{\textbf{u}}{\text{maximize}} \quad \sum_{k=1}^{K_t} R_k(\textbf{u}) \nonumber \\
	&\text{subject to} \quad  
	\tilde{R}_k(\textbf{u}, \tilde{\gamma}_k)  \ge \frac{D_s}{\delta_k} , \quad  k \in \mathcal{K}_s. \label{eq:problem_re}
\end{align}
where $\delta_k$ is the communication latency requirement of user $k$.
Note that for tractability, the upper bound of the communication latency is used in the first constraint. This is reasonable since satisfying the latency requirement with the upper bound latency always guarantees that with the actual latency.
We assume that the \ac{BS} knows the predefined latency requirement of delay-constrained users.
However, since the optimization problem is not convex, it is difficult to obtain the optimal solution for \eqref{eq:problem_re}.
Instead of that, we can obtain the sub-optimal solution by checking the first-order optimality conditions.

\subsection{Local Optimal Condition}
In the following Theorem \ref{theorm:first-order}, we present the first-order optimality condition of the optimization problem in \eqref{eq:problem_re} for the precoding vector and the Lagrangian multiplier.
\begin{theorem} \label{theorm:first-order}
	Let $\phi(\textbf{u},\boldsymbol{\lambda})=\frac{\prod_{k=1}^{K_t} \textbf{u}^\text{H} \textbf{A}_k \textbf{u}
	\prod_{k=K_t+1}^{K_t+K_s} \left(\textbf{u}^\text{H} \textbf{A}_k \textbf{u}\right)^{\lambda_k \left(1-f_k\left(\tilde{\gamma}_k\right)\right)}} {\prod_{k=1}^{K_t} \textbf{u}^\text{H} \textbf{B}_k \textbf{u} 
	\prod_{k=K_t+1}^{K_t+K_s} \left(\textbf{u}^\text{H} \textbf{B}_k \textbf{u}\right)^{\lambda_k \left(1-f_k\left(\tilde{\gamma}_k\right)\right)}}$
	for the Lagrangian multiplier of user $k$, $\lambda_k$.
	The first-order optimality condition satisfies when
	\begin{align}
	\bar{\textbf{A}}\left(\textbf{u},\boldsymbol{\lambda}\right) \textbf{u}
	= \phi(\textbf{u},\boldsymbol{\lambda}) \bar{\textbf{B}}\left(\textbf{u},\boldsymbol{\lambda}\right) \textbf{u} \label{eq:KKT_first}
	\end{align}
	where $\bar{\textbf{A}}\left(\textbf{u},\boldsymbol{\lambda}\right)$ and $\bar{\textbf{B}}\left(\textbf{u},\boldsymbol{\lambda}\right)$ are given by
	\begin{align}
	\bar{\textbf{A}}\left(\textbf{u},\boldsymbol{\lambda}\right) &\hspace{-0.7mm}=\hspace{-0.7mm}
	\prod_{k=1}^{K_t} \hspace{-0.3mm} \textbf{u}^\text{H} \hspace{-0.3mm} \textbf{A}_k \textbf{u}
	\hspace{-0.3mm} \prod_{k=K_t+1}^{K_t+K_s} \hspace{-1mm} \left(\textbf{u}^\text{H} \hspace{-0.3mm} \textbf{A}_k \textbf{u}\right)^{\hspace{-0.3mm} \lambda_k \left(1-f_k\left(\tilde{\gamma}_k\right)\right) } 
	\hspace{-0.7mm} \left\{
	\sum_{k=1}^{K_t} \frac{2 \textbf{A}_k} {\textbf{u}^\text{H} \hspace{-0.3mm} \textbf{A}_k \textbf{u}}
	\hspace{-0.7mm} + \hspace{-1mm}
	\sum_{k=K_t+1}^{K_t+K_s} \frac{2 \lambda_k \hspace{-0.7mm} \left(1 \hspace{-0.7mm} - \hspace{-0.7mm} f_k\hspace{-0.7mm}\left(\tilde{\gamma}_k\right)\right) \hspace{-0.3mm} \textbf{A}_k} {\textbf{u}^\text{H} \textbf{A}_k \textbf{u}} \hspace{-0.3mm}
	\right\} \hspace{-0.5mm}, \hspace{-1.5mm} \label{eq:bar_A} \\
	\bar{\textbf{B}}\left(\textbf{u},\boldsymbol{\lambda}\right) &\hspace{-0.7mm}=\hspace{-0.7mm}
	\prod_{k=1}^{K_t} \hspace{-0.3mm} \textbf{u}^\text{H} \textbf{B}_k \textbf{u}
	\hspace{-0.3mm}\prod_{k=K_t+1}^{K_t+K_s} \hspace{-1mm} \left(\textbf{u}^\text{H} \textbf{B}_k \textbf{u}\right)^{\hspace{-0.3mm} \lambda_k \left(1-f_k\left(\tilde{\gamma}_k\right)\right) }
	\hspace{-0.7mm}\left\{
	\sum_{k=1}^{K_t} \frac{2 \textbf{B}_k} {\textbf{u}^\text{H} \textbf{B}_k \textbf{u}}
	\hspace{-0.7mm} + \hspace{-1mm}  \sum_{k=K_t+1}^{K_t+K_s} \frac{2 \lambda_k \hspace{-0.7mm} \left(1 \hspace{-0.8mm} - \hspace{-0.8mm} f_k\hspace{-0.7mm} \left(\tilde{\gamma}_k\right)\right) \textbf{B}_k} {\textbf{u}^\text{H} \textbf{B}_k \textbf{u}} \hspace{-0.3mm}
	\right\} \hspace{-0.5mm}. \hspace{-1.5mm} \label{eq:bar_B}
	\end{align}
	The Lagrangian multipliers $\boldsymbol{\lambda}$ are chosen so that $\textbf{u}$ satisfies
	\begin{align} \label{eq:lagrange}
	\log_2 \left\{  \left(\frac{\textbf{u}^\text{H} \textbf{A}_k \textbf{u}} {\textbf{u}^\text{H} \textbf{B}_k \textbf{u}}\right)^{1 - f_k\left(\tilde{\gamma}_k\right)} \right\} 
	- g_k\left(\tilde{\gamma}_k\right)  = \frac{D_s}{\delta_k} , \quad  k \in \mathcal{K}_s.
	\end{align}
	If the equation \eqref{eq:lagrange} cannot be satisfied, $\boldsymbol{\lambda}$ are chosen so that $\textbf{u}$ satisfies $\tilde{R}_k(\textbf{u}, \tilde{\gamma}_k)  > \frac{D_s}{\delta_k} , \,  k \in \mathcal{K}_s$.
\end{theorem}
\begin{IEEEproof}
	From \eqref{eq:rate_net}, \eqref{eq:latency}, and \eqref{eq:problem_re}, we can define the Lagrangian function as
	\begin{align}
	\mathcal{L}(\textbf{u}, \boldsymbol{\lambda}) &=
	- \sum_{k=1}^{K_t} \log_2  \left(\frac{\textbf{u}^\text{H} \textbf{A}_k \textbf{u}} {\textbf{u}^\text{H} \textbf{B}_k \textbf{u}}\right)
	- \sum_{k=K_t+1}^{K_t+K_s} \lambda_k \left[ \log_2 \left\{  \left(\frac{\textbf{u}^\text{H} \textbf{A}_k \textbf{u}} {\textbf{u}^\text{H} \textbf{B}_k \textbf{u}}\right)^{1 - f_k\left(\tilde{\gamma}_k\right)} \right\} - g_k\left(\tilde{\gamma}_k\right) - \frac{D_s}{\delta_k} \right] \nonumber \\
	&= - \log_2 \left\{\prod_{k=1}^{K_t} \frac{\textbf{u}^\text{H} \textbf{A}_k \textbf{u}} {\textbf{u}^\text{H} \textbf{B}_k \textbf{u}}  \right\}
	- \log_2 \left\{ \prod_{k=K_t+1}^{K_t+K_s} \left(\frac{\textbf{u}^\text{H} \textbf{A}_k \textbf{u}} {\textbf{u}^\text{H} \textbf{B}_k \textbf{u}}\right)^{ \lambda_k \left(1 - f_k\left(\tilde{\gamma}_k\right) \right) } \right\}  \nonumber \\
	& \quad + \log_2 \left\{ \prod_{k=K_t+1}^{K_t+K_s} 2^{\lambda_k g_k\left(\tilde{\gamma}_k\right) } \right\}
	+ \sum_{k=K_t+1}^{K_t+K_s} \frac{\lambda_k D_s}{\delta_k} \nonumber \\
	& = -\log_2 \phi(\textbf{u}, \boldsymbol{\lambda}) + \log_2 \left\{ \prod_{k=K_t+1}^{K_t+K_s} 2^{\lambda_k g_k\left(\tilde{\gamma}_k\right) } \right\}
	+ \sum_{k=K_t+1}^{K_t+K_s} \frac{\lambda_k D_s}{\delta_k} . \label{eq:Lagrangian}
	\end{align}
	From \eqref{eq:Lagrangian}, we first obtain $\frac{\partial \mathcal{L}(\textbf{u}, \boldsymbol{\lambda})} {\partial \textbf{u}^\text{H}}$ as
	\begin{align}
	\frac{\partial \mathcal{L}(\textbf{u}, \boldsymbol{\lambda})} {\partial \textbf{u}^\text{H}}
	=
	-\frac{1} {\phi(\textbf{u}, \boldsymbol{\lambda}) \ln 2}
	\frac{\partial \phi(\textbf{u}, \boldsymbol{\lambda})} {\partial \textbf{u}^\text{H}} . \label{eq:der_Lag}
	\end{align}
	In \eqref{eq:der_Lag}, since $\phi(\textbf{u}, \boldsymbol{\lambda}) > 0$, $\frac{\partial \mathcal{L}(\textbf{u}, \boldsymbol{\lambda})} {\partial \textbf{u}} = 0$ is equivalent to $\frac{\partial \phi(\textbf{u}, \boldsymbol{\lambda})} {\partial \textbf{u}^\text{H}} = 0$, which can be presented as
	\begin{align}
	\frac{\partial \phi(\textbf{u}, \boldsymbol{\lambda})} {\partial \textbf{u}^\text{H}}=0 
	&\iff \phi(\textbf{u}, \boldsymbol{\lambda}) 
	\left\{
	\frac{\nabla_{\textbf{u}^\text{H}} \phi_1(\textbf{u})} {\phi_1(\textbf{u})}
	+ \frac{\nabla_{\textbf{u}^\text{H}} \phi_2(\textbf{u}, \boldsymbol{\lambda})} {\phi_2(\textbf{u}, \boldsymbol{\lambda})}
	- \frac{\nabla_{\textbf{u}^\text{H}} \phi_3(\textbf{u})} {\phi_3(\textbf{u})}
	- \frac{\nabla_{\textbf{u}^\text{H}} \phi_4(\textbf{u}, \boldsymbol{\lambda})} {\phi_4(\textbf{u}, \boldsymbol{\lambda})}
	\right\} = 0 \nonumber \\
	&\iff \phi(\textbf{u}, \boldsymbol{\lambda}) 
	\left[
	\sum_{k=1}^{K_t} \frac{2 \textbf{A}_k \textbf{u}} {\textbf{u}^\text{H} \textbf{A}_k \textbf{u}}
	+ \sum_{k=K_t+1}^{K_t+K_s} \frac{2 \lambda_k \left(1 - f_k\left(\tilde{\gamma}_k\right) \right) \textbf{A}_k \textbf{u}} {\textbf{u}^\text{H} \textbf{A}_k \textbf{u}}
	- \sum_{k=1}^{K_t} \frac{2 \textbf{B}_k \textbf{u}} {\textbf{u}^\text{H} \textbf{B}_k \textbf{u}}
	\right. \nonumber \\
	& \left. \quad \quad \quad \quad \quad \quad \,\,\,
	- \sum_{k=K_t+1}^{K_t+K_s} \frac{2 \lambda_k \left(1 - f_k\left(\tilde{\gamma}_k\right) \right) \textbf{B}_k \textbf{u}} {\textbf{u}^\text{H} \textbf{B}_k \textbf{u}}
	\right] = 0 \label{eq:der_phi}
	\end{align}
	where $\phi_1(\textbf{u})$, $\phi_2(\textbf{u}, \boldsymbol{\lambda})$, $\phi_3(\textbf{u})$, and $\phi_4(\textbf{u}, \boldsymbol{\lambda})$ are given by
	\begin{align}
	&\phi_1(\textbf{u}) = 
	\prod_{k=1}^{K_t} \textbf{u}^\text{H} \textbf{A}_k \textbf{u} , \quad
	\phi_2(\textbf{u}, \boldsymbol{\lambda}) = 
	\prod_{k=K_t+1}^{K_t+K_s} \left(\textbf{u}^\text{H} \textbf{A}_k \textbf{u}\right)^{\lambda_k \left(1 - f_k\left(\tilde{\gamma}_k\right)\right) }, \nonumber \\
	&
	\phi_3(\textbf{u}) =
	\prod_{k=1}^{K_t} \textbf{u}^\text{H} \textbf{B}_k \textbf{u} , \quad
	\phi_4(\textbf{u}, \boldsymbol{\lambda}) = 
	\prod_{k=K_t+1}^{K_t+K_s} \left(\textbf{u}^\text{H} \textbf{B}_k \textbf{u}\right)^{\lambda_k \left(1 - f_k\left(\tilde{\gamma}_k\right)\right) } .
	\end{align}
	From \eqref{eq:der_phi}, the first-order optimality condition with respect to $\textbf{u}$, i.e., $\frac{\partial \mathcal{L}(\textbf{u}, \boldsymbol{\lambda})} {\partial \textbf{u}}=0$, can be presented as \eqref{eq:KKT_first}.
	
	In addition, by setting the first derivative of $\mathcal{L}(\textbf{u}, \boldsymbol{\lambda})$ with respect to $\lambda_k$ to zero, i.e., $\frac{\partial \mathcal{L}(\textbf{u}, \boldsymbol{\lambda})} {\partial \lambda_k}=0$, we can readily obtain the first-order optimality condition with respect to $\lambda_k$ as \eqref{eq:lagrange}.
\end{IEEEproof}

From Theorem \ref{theorm:first-order}, we can obtain the first-order optimality condition for the precoding vector, expressed as the generalized eigenvalue problem in \eqref{eq:KKT_first}.
It means that any stationary point of the problem in \eqref{eq:problem_re} is the eigenvector of the matrix $\bar{\textbf{B}}\left(\textbf{u},\boldsymbol{\lambda}\right)^{-1} \bar{\textbf{A}}\left(\textbf{u},\boldsymbol{\lambda}\right)$.
Therefore, when we interpret $\phi(\textbf{u},\boldsymbol{\lambda})$ as the eigenvalue of the matrix $\bar{\textbf{B}}\left(\textbf{u},\boldsymbol{\lambda}\right)^{-1} \bar{\textbf{A}}\left(\textbf{u},\boldsymbol{\lambda}\right)$ and treat $\textbf{u}$ as the eigenvector corresponding to the eigenvalue, we can obtain an optimal solution of the problem in \eqref{eq:problem_re} by finding a principal eigenvector of the matrix $\bar{\textbf{B}}\left(\textbf{u},\boldsymbol{\lambda}\right)^{-1} \bar{\textbf{A}}\left(\textbf{u},\boldsymbol{\lambda}\right)$.
Although the conventional generalized eigenvalue problem can be solved, we cannot solve a generalized eigenvalue problem in \eqref{eq:KKT_first} as the matrix $\bar{\textbf{B}}\left(\textbf{u},\boldsymbol{\lambda}\right)^{-1} \bar{\textbf{A}}\left(\textbf{u},\boldsymbol{\lambda}\right)$ is also affected by $\textbf{u}$.
In spite of this difficulty, using the power iteration algorithm \cite{Gol:96}, we provide a computationally efficient algorithm that finds a converged feasible precoding vector that satisfies the first-order optimality condition in \eqref{eq:KKT_first}.

\subsection{Precoding Algorithm}
\begin{algorithm} [t]
	\caption{Delay-GPI} \label{alg:Algorithm1} 
	\begin{algorithmic} 
		\State {\bf{initialize}}: $\textbf{u}^{(0)}=\text{RZF}$, $\textbf{u}^{(-1)}=\textbf{0}$, $\lambda_k^{(0)}$, and $\xi$.
		\State Set the iteration count $n = 0$ and $j = 0$.
		%\State {(\bf{The precoding update phase})}
		\WHILE {$\tilde{R}_k(\textbf{u}^{(j)}, \tilde{\gamma}_k) < \frac{D_s}{\delta_k}, \quad \forall k \in \mathcal{K}_s$}
		\State $n \leftarrow n+1$.
		\State $\lambda_k^{(n)} \leftarrow \left[\lambda_k^{(n-1)} + \Delta\lambda_k^{(n)} \right]^+ , \quad \forall k \in \mathcal{K}_s$.
		\WHILE {$\left\|\textbf{u}^{(j)} - \textbf{u}^{(j-1)} \right\| > \xi$}
		\State $j \leftarrow j+1$.
		\State Create the matrices $\bar{\textbf{A}} \left(\textbf{u}^{(j-1)},\boldsymbol{\lambda}^{(n)}\right)$ and $\bar{\textbf{B}} \left(\textbf{u}^{(j-1)},\boldsymbol{\lambda}^{(n)}\right)$ by using \eqref{eq:bar_A} and \eqref{eq:bar_B}. 
		\State Update $\textbf{u}^{(j)} = \left[\bar{\textbf{B}} \left(\textbf{u}^{(j-1)},\boldsymbol{\lambda}^{(n)}\right)\right]^{-1} \bar{\textbf{A}} \left(\textbf{u}^{(j-1)},\boldsymbol{\lambda}^{(n)}\right) \textbf{u}^{(j-1)}$.
		\State {Normalize $\textbf{u}^{(j)} = \textbf{u}^{(j)}/\left\| \textbf{u}^{(j)}\right\|$}. 
		%\State Update {$R_{{\rm e}, m} $} as in \eqref{eq:wiretap_se} with $\bar {\bf{f}}^{(t)}$ for $m \in \CMcal{M}$
		%\State Update {$w_m $} as in \eqref{eq:weight}
		\ENDWHILE
		\ENDWHILE
		%\State {\bf{output}}: $\lfloor \left(b_n^{\star}\right)^+ \rfloor$ for $n \in \{1,...,N\}$
		\State {\bf{output}}: $\textbf{u}^{\star} = \textbf{u}^{(j)}$
	\end{algorithmic}
\end{algorithm}
In this subsection, we develop an algorithm that finds the principal eigenvector of a generalized eigenvalue problem in \eqref{eq:KKT_first} as presented in Algorithm \ref{alg:Algorithm1}.
We denote a proposed algorithm as the generalized power iteration with the latency constraint (Delay-GPI).
In the proposed algorithm, we iteratively obtain an optimal precoding vector $\textbf{u}$.
In the $j$-th iteration, for given $\textbf{u}^{(j-1)}$ and $\boldsymbol{\lambda}^{(n)}$, we construct the matrices $\bar{\textbf{A}} \left(\textbf{u}^{(j-1)},\boldsymbol{\lambda}^{(n)}\right)$ and $\bar{\textbf{B}} \left(\textbf{u}^{(j-1)},\boldsymbol{\lambda}^{(n)}\right)$ using Theorem \ref{theorm:first-order}.
We then update the precoding vector $\textbf{u}^{(j)} = \left[\bar{\textbf{B}} \left(\textbf{u}^{(j-1)},\boldsymbol{\lambda}^{(n)}\right)\right]^{-1} \bar{\textbf{A}} \left(\textbf{u}^{(j-1)},\boldsymbol{\lambda}^{(n)}\right) \textbf{u}^{(j-1)}$ by using the \ac{GPI} method \cite{Cho:20} and normalize it as $\textbf{u}^{(j)} = \textbf{u}^{(j)}/\left\| \textbf{u}^{(j)}\right\|$.
Until the precoding vector converges to the principal eigenvector, i.e., $\left\|\textbf{u}^{(j)} - \textbf{u}^{(j-1)} \right\| < \xi$ with a predetermined tolerance parameter $\xi$, we repeat this process.
From the convergent precoding vector $\textbf{u}^{(j)}$, we check the latency requirements of delay-constrained users.
If latency requirements are satisfied, i.e., $\tilde{R}_k(\textbf{u}^{(j)}, \tilde{\gamma}_k) \ge \frac{D_s}{\delta_k}  , \, \forall k \in \mathcal{K}_s$, then the algorithm ends. Otherwise, we update Lagrangian multipliers to satisfy latency requirements.

\begin{remark}
	(Precoding vector feasibility)
	For a given latency requirement $\delta_k$, according to network parameters (e.g., decoding error probability $\varepsilon_k$ and data size of transmit symbol $D_s$), the Delay-GPI cannot converge to a principal eigenvector.
	This is because when the channel gain of a delay-constrained user is small, the BS should transmit the downlink signal to the user with larger power.
	However, due to the transmit power constraint, the latency requirement may not be satisfied.
	Hence, depending on the channel gain and the latency requirement, we should use enough average transmit power $P$ to make the precoding vector design feasible.
\end{remark}
\begin{remark}
	(Algorithm complexity)
	As we use the GPI-based algorithm, the complexity of the Delay-GPI algorithm is determined by the computing for $\left[\bar{\textbf{B}} \left(\textbf{u}^{(j-1)},\boldsymbol{\lambda}^{(n)}\right)\right]^{-1}$ \cite{Cho:20}.
	In our problem, the matrix $\left[\bar{\textbf{B}} \left(\textbf{u}^{(j-1)},\boldsymbol{\lambda}^{(n)}\right)\right]$ is a linear combination of $\textbf{B}_k \left(\textbf{u}^{(j-1)}\right)$, which is a block diagonal and symmetric matrix.
	Therefore, the algorithm requires the computational complexity order of $\mathcal{O}\left(\frac{1}{3} \left(K_t+K_s\right) N^3\right)$ to obtain $\left[\bar{\textbf{B}} \left(\textbf{u}^{(j-1)},\boldsymbol{\lambda}^{(n)}\right)\right]^{-1}$.
	Furthermore, to satisfy the latency requirement and converge the precoding vector, we need a total computational complexity order of $\mathcal{O}\left(\frac{1}{3} N_c N_p \left(K_t+K_s\right) N^3\right)$ where $N_c$ is the number of iterations for the latency constraint and $N_p$ is the number of iterations for the precoding vector convergence.
\end{remark}
\begin{remark}
	(Imperfect \ac{CSIT}) In this paper, we assume that the \ac{BS} has the perfect knowledge of \ac{CSI}.
	However, due to the imperfect channel estimation and the pilot contamination, the \ac{BS} can have limited knowledge of \ac{CSI}.
	Even under such circumstances, by using the generalized mutual information, we can obtain a lower bound of the ergodic spectral efficiency as \cite{Cho:20}
	\begin{align}
		R_k^\text{imp}(\textbf{u}) 
		&= \log_2 \hspace{-1.3mm} \left(\frac{\sum_{i=1}^{K_t + K_s} \hspace{-0.5mm} \textbf{u}_i^\text{H} \hspace{-0.8mm} \left(\hat{\textbf{h}}_k \hat{\textbf{h}}_k^\text{H}\right) \hspace{-0.5mm} \textbf{u}_i 
		\hspace{-0.8mm} + \hspace{-0.8mm}
		 \sum_{i=1}^{K_t + K_s} \hspace{-0.6mm} \textbf{u}_i^\text{H} \boldsymbol{\Phi}_k \textbf{u}_i 
		 \hspace{-0.8mm} + \hspace{-0.8mm}
		  \frac{\sigma_k^2}{P}}
		{\sum_{i \neq k}^{K_t + K_s} \hspace{-0.5mm} \textbf{u}_i^\text{H} \hspace{-0.8mm} \left(\hat{\textbf{h}}_k \hat{\textbf{h}}_k^\text{H}\right) \hspace{-0.5mm} \textbf{u}_i 
		\hspace{-0.8mm} + \hspace{-0.8mm}
		\sum_{i=1}^{K_t + K_s} \hspace{-0.6mm} \textbf{u}_i^\text{H} \boldsymbol{\Phi}_k \textbf{u}_i 
		\hspace{-0.8mm} + \hspace{-0.8mm}
		\frac{\sigma_k^2}{P}} \hspace{-0.5mm} \right) 
		\hspace{-1.5mm} = \hspace{-0.7mm} \log_2 \hspace{-1.3mm} \left(\hspace{-0.3mm} \frac{\textbf{u}^\text{H} \textbf{A}_k^\text{imp} \textbf{u}} {\textbf{u}^\text{H} \textbf{B}_k^\text{imp} \textbf{u}} \hspace{-0.5mm}\right) \hspace{-0.3mm} , \quad k \hspace{-0.6mm} \in \hspace{-0.6mm} \mathcal{K}_t , \nonumber \\
		\tilde{R}_k^\text{imp}(\textbf{u}, \tilde{\gamma}_k)
		&= \log_2 \left\{  \left(\frac{\textbf{u}^\text{H} \textbf{A}_k^\text{imp} \textbf{u}} {\textbf{u}^\text{H} \textbf{B}_k^\text{imp} \textbf{u}}\right)^{1 - f_k\left(\tilde{\gamma}_k\right)} \right\} - g_k\left(\tilde{\gamma}_k\right) , \quad  k \in \mathcal{K}_s
		\label{eq:rate_error_lower}
	\end{align}
	where $\hat{\textbf{h}}_k = \textbf{h}_k - \textbf{e}_k$ and $\textbf{e}_k = \left[e_k^1, e_k^2, \cdots, e_k^N\right] \in \mathbb{C}^{N \times 1}$ is the channel estimation error vector, randomly distributed as the complex Gaussian, i.e., $\textbf{e}_k \hspace{-0.6mm} \sim \hspace{-0.6mm} \mathcal{CN}(0,\boldsymbol{\Phi}_k)$ where $\boldsymbol{\Phi}_k \hspace{-0.6mm} = \hspace{-0.6mm} \mathbb{E} \hspace{-0.6mm}\left[\textbf{e}_k \textbf{e}_k^\text{H}\right] \hspace{-0.6mm} \in \hspace{-0.6mm} \mathbb{C}^{N \times N}$ is the covariance matrix of the estimation error. In \eqref{eq:rate_error_lower}, $\textbf{A}_k^\text{imp} \hspace{-0.5mm} = \hspace{-0.5mm} \text{diag} \hspace{-0.6mm} \left(\hat{\textbf{h}}_k \hat{\textbf{h}}_k^\text{H} \hspace{-0.4mm} + \hspace{-0.4mm} \boldsymbol{\Phi}_k, \cdots ,\hat{\textbf{h}}_k \hat{\textbf{h}}_k^\text{H} \hspace{-0.4mm} + \hspace{-0.4mm} \boldsymbol{\Phi}_k\right) + \frac{\sigma_k^2}{P} \textbf{I}_{N(K_t+K_s)}$ and $\textbf{B}_k^\text{imp} = \textbf{A}_k^\text{imp} - \text{diag} \left(\textbf{0},\cdots,\hat{\textbf{h}}_k \hat{\textbf{h}}_k^\text{H},\cdots,\textbf{0}\right)$.
	By using \eqref{eq:rate_error_lower} in \eqref{eq:problem_re}, we can readily obtain the first-order optimality condition and present a sub-optimal solution of the imperfect \ac{CSIT} case using the Delay-GPI.
	%provides lower performance than perfect \ac{CSIT} and
\end{remark}

\section{Precoding Design for DCTU-MIMO network with \ac{IR-HARQ} scheme} \label{sec:Algorithm_IR-HARQ}
In this section, we formulate optimization problems that maximize the spectral efficiency of delay-tolerant users while satisfying the communication latency constraint of delay-constrained users for a DCTU-MIMO network with the \ac{IR-HARQ} scheme.
We then derive the first-order optimality condition of optimization problems and extend the proposed Delay-GPI for the design of precoding vectors in a DCTU-MIMO network with the \ac{IR-HARQ} scheme.

\subsection{Problem Formulation}
Using \eqref{eq:rate_HARQ}, \eqref{eq:rate_s_HARQ_bound}, and \eqref{eq:rate_approx}, we formulate the optimization problem for a DCTU-MIMO network with the IR-HARQ scheme at the first transmission round as
\begin{align}
&\underset{\textbf{u}_\text{IR}}{\text{maximize}} \quad 
\sum_{k=1}^{K_t} R_{k,1}(\textbf{u}_\text{IR}) \nonumber \\
& \text{subject to} \quad  
\tilde{R}_{k,1}(\textbf{u}_\text{IR}, \tilde{\gamma}_k(1))  
\ge \frac{D_s}{\delta_k}
- \sum_{t=2}^{T} \hat{R}_{k,t}(\textbf{u}_\text{IR}, \tilde{\gamma}_k(t)) , \quad  k \in \mathcal{K}_s . \label{eq:problem_HARQ}
\end{align}
Furthermore, the optimization problem for a DCTU-MIMO network with the IR-HARQ at the $t_o$-th transmission round ($2 \le t_o < T$) can be formulated as

\begin{align}
&\underset{\textbf{u}_{\text{IR},t_o}}{\text{maximize}} \quad 
\sum_{k=1}^{K_t} R_{k,t_o}\left(\textbf{u}_{\text{IR},t_o}\right) \nonumber \\
& \text{subject to} \quad  
\tilde{R}_{k,t_o}\left(\textbf{u}_{\text{IR}, t_o}, \tilde{\gamma}_k(t_o)\right)  
\ge \frac{D_s}{\delta_k}
- \sum_{t=t_o+1}^{T} \hat{R}_{k,t}\left(\textbf{u}_{\text{IR}, t_o}, \tilde{\gamma}_k(t)\right)
- \sum_{t=1}^{t_o-1} \bar{R}_k(t) , \quad  k \in \mathcal{K}_s \label{eq:problem_HARQ_sec}
\end{align}
where $\textbf{u}_{\text{IR},t_o} \hspace{-1mm} = \hspace{-1mm} \left[\textbf{u}_\text{IR}^\text{H}\hspace{-0.4mm}(t_o),\cdots,\textbf{u}_\text{IR}^\text{H}\hspace{-0.4mm}(T)\right]^\text{H}$, $\textbf{u}_{\text{IR},1} \hspace{-1mm} = \hspace{-1mm} \textbf{u}_\text{IR}$, $\bar{R}_k\hspace{-0.4mm}(t) \hspace{-1mm} = \hspace{-1mm} \tilde{R}_{k,t}\hspace{-0.4mm}\left(\textbf{u}_{\text{IR},t}, \tilde{\gamma}_k\hspace{-0.4mm}(t)\right)$ is the spectral efficiency of the delay-constrained user obtained at the $t$-th transmission round, and $\tilde{R}_{k,t_o}\left(\textbf{u}_{\text{IR},t_o}, \tilde{\gamma}_k(t_o)\right)$ and $\hat{R}_{k,t}\left(\textbf{u}_{\text{IR},t_o}, \tilde{\gamma}_k(t)\right)$ are the spectral efficiency of the delay-constrained user at the $t_o$-th transmission round and the approximation of the ergodic spectral efficiency of the delay-constrained user at the $t$-th transmission round ($t_o+1 \le t \le T$), respectively, given by
\begin{align}
	\tilde{R}_{k,t_o}\left(\textbf{u}_{\text{IR},t_o}, \tilde{\gamma}_k(t_o)\right)
	&= 
	\log_2 \left\{ 
	\left(\frac{\textbf{u}_{\text{IR},t_o}^\text{H} \textbf{D}_{k,t_o}(t_o) \textbf{u}_{\text{IR},t_o}} {\textbf{u}_{\text{IR},t_o}^\text{H} \textbf{E}_{k,t_o}(t_o) \textbf{u}_{\text{IR},t_o}}\right)^{1-f_{k,t_o}\left(\tilde{\gamma}_k(t_o)\right)} \right\}  
	- g_{k,t_o}\left(\tilde{\gamma}_k(t_o)\right) , \,\, k \in \mathcal{K}_s , \nonumber \\
	\hat{R}_{k,t}\left(\textbf{u}_{\text{IR},t_o}, \tilde{\gamma}_k(t)\right)
	&= 
	\log_2 \left\{ 
	\left(\frac{\textbf{u}_{\text{IR},t_o}^\text{H} \tilde{\textbf{D}}_{k,t}(t) \textbf{u}_{\text{IR},t_o}} {\textbf{u}_{\text{IR},t_o}^\text{H} \tilde{\textbf{E}}_{k,t}(t) \textbf{u}_{\text{IR},t_o}}\right)^{1-f_{k,t}\left(\tilde{\gamma}_k(t)\right)} \right\}  
	- g_{k,t}\left(\tilde{\gamma}_k(t)\right) , \,\, k \in \mathcal{K}_s . \label{eq:rate_middle}
\end{align}
In \eqref{eq:rate_middle}, $\textbf{D}_{k,t_o}(t_o)$ and $\tilde{\textbf{D}}_{k,t}(t)$ are given by
	\begin{align}
		\textbf{D}_{k,t_o}\hspace{-0.5mm}(\hspace{-0.2mm}t_o\hspace{-0.3mm}) \hspace{-0.8mm} &= \hspace{-1.5mm} \begin{bmatrix}
			\textbf{D}_{k,t_o}^{t_o}(t_o) & \hspace{-1mm} \textbf{0}  & \hspace{-1mm} \dots & \hspace{-1mm}   \textbf{0}  \vspace{-1mm} \\
			\textbf{0} & \hspace{-1mm} \textbf{0}  & \hspace{-1mm} \dots & \hspace{-1mm} \textbf{0} \vspace{-1mm} \\
			\vdots   & \hspace{-1mm} \vdots  & \hspace{-1mm} \ddots  & \hspace{-1mm} \vdots \vspace{-1mm} \\   
			\textbf{0} & \hspace{-1mm} \textbf{0} & \hspace{-1mm} \dots & \hspace{-1mm} \textbf{0}   \\
		\end{bmatrix}
		\hspace{-1.2mm} + \hspace{-0.7mm}
		\frac{\sigma_k^2}{P} \textbf{I}_{N(K_t+K_s)(T-t_o+1)} \hspace{-0.8mm} \in \hspace{-0.8mm} \mathbb{C}^{N(K_t+K_s)(T-t_o+1) \times N(K_t+K_s)(T-t_o+1)} , \hspace{-2mm} \nonumber \\
		\tilde{\textbf{D}}_{k,t}\hspace{-0.4mm}(t) \hspace{-0.8mm} &= \hspace{-1.5mm} \begin{bmatrix}
			\textbf{0} & \hspace{-2mm} \dots & \hspace{-2mm} \textbf{0} & \hspace{-2mm} \dots & \hspace{-2mm}   \textbf{0}    \vspace{-1mm}   \\
			\vdots  & \hspace{-2mm} \ddots & \hspace{-2mm} \vdots & \hspace{-2mm}     & \hspace{-2mm}  \vdots \vspace{-1mm} \\
			\textbf{0} & \hspace{-2mm} \dots & \hspace{-2mm}   \tilde{\textbf{D}}_{k,t}^t\hspace{-0.4mm}(t) & \hspace{-2mm} \dots & \hspace{-2mm} \textbf{0}   \vspace{-1mm}    \\
			\vdots  &\hspace{-2mm}     &  \hspace{-2mm}  \vdots & \hspace{-2mm} \ddots  & \hspace{-2mm}  \vdots \vspace{-1mm}  \\   
			\textbf{0}   & \hspace{-2mm} \dots   & \hspace{-2mm} \textbf{0} & \hspace{-2mm} \dots   & \hspace{-2mm}   \textbf{0}      \\
		\end{bmatrix}
		\hspace{-1.5mm} + \hspace{-0.7mm}
		\frac{\sigma_k^2}{P} \textbf{I}_{N\hspace{-0.3mm}(K_t+K_s)(T-t_o+1)} \hspace{-1mm} \in \hspace{-0.8mm} \mathbb{C}^{N\hspace{-0.3mm}(K_t+K_s)(T-t_o+1) \hspace{-0.3mm} \times \hspace{-0.3mm} N(K_t+K_s)(T-t_o+1)}  \hspace{-1.5mm} \label{eq:matrix_D_middle}
	\end{align}
	where $\textbf{D}_{k,t_o}^{t_o}(t_o) \hspace{-0.7mm} = \hspace{-0.7mm} \text{diag} \left(\textbf{h}_k(t_o) \textbf{h}_k^\text{H}(t_o) , \cdots, \textbf{h}_k(t_o) \textbf{h}_k^\text{H}(t_o)\right) \hspace{-0.7mm} \in \hspace{-0.7mm} \mathbb{C}^{N(K_t+K_s) \times N(K_t+K_s)}$ and $\tilde{\textbf{D}}_{k,t}^{t}(t) \hspace{-0.7mm} = \hspace{-0.7mm} \text{diag} \left(\textbf{C}_k(t) , \cdots, \textbf{C}_k(t) \right) \hspace{-0.7mm} \in \hspace{-0.7mm} \mathbb{C}^{N(K_t+K_s) \times N(K_t+K_s)}$ is $t$-th sub-block matrix of $\tilde{\textbf{D}}_{k,t}(t)$.
	In addition, $\textbf{E}_{k,t_o}(t_o) = \textbf{D}_{k,t_o}(t_o) - \text{diag} \left(\textbf{0},\cdots,\textbf{h}_k(t_o) \textbf{h}_k^\text{H}(t_o),\cdots,\textbf{0}\right)$ is constructed by subtracting the $k$-th sub-block matrix from $\textbf{D}_{k,t_o}^{t_o}(t_o)$ and $\tilde{\textbf{E}}_{k,t}(t) = \tilde{\textbf{D}}_{k,t}(t) - \text{diag} \left(\textbf{0},\cdots,\textbf{C}_k(t),\cdots,\textbf{0}\right)$ is constructed by subtracting the $k$-th sub-block matrix from $\tilde{\textbf{D}}_{k,t}^{t}(t)$.
%In the $t_o$-th transmission round, we note that optimize the precoding vector except for the corresponding precoding vectors up to the $t_o-1$ transmission round because we obtained the optimal precoding vector until the $t_o-1$ transmission round.
After $T-1$ transmission rounds, the optimization problem for a DCTU-MIMO network with the IR-HARQ at the last transmission round is expressed as

\begin{align}
&\underset{\textbf{u}_{\text{IR},T}}{\text{maximize}} \quad 
\sum_{k=1}^{K_t} R_{k,T}\left(\textbf{u}_{\text{IR},T}\right) \nonumber \\
& \text{subject to} \quad  
\tilde{R}_{k,T}\left(\textbf{u}_{\text{IR},T}, \tilde{\gamma}_k(T)\right)
\ge \frac{D_s}{\delta_k} 
- \sum_{t=1}^{T-1} \bar{R}_k(t) , \quad  k \in \mathcal{K}_s \label{eq:problem_HARQ_Last}
\end{align}
where $\textbf{u}_{\text{IR},T} = \left[\textbf{u}_1^\text{H}(T),\cdots,\textbf{u}_{K_t+K_s}^\text{H}(T)\right]^\text{H}$.
However, the optimization problems in \eqref{eq:problem_HARQ}, \eqref{eq:problem_HARQ_sec}, and \eqref{eq:problem_HARQ_Last} are not convex, so it is difficult to obtain the optimal solutions.
Hence, we obtain the sub-optimal solution by checking the first-order optimality conditions.

\subsection{Local Optimal Condition}
In the following Theorem \ref{theorm:first-order_HARQ}, we present the first-order optimality conditions of the optimization problem at the $t_o$-th transmission round ($2 \le t_o <T$) for the precoding vector and the Lagrangian multiplier.
Let us define
$\phi^\text{IR}\left(\textbf{u}_{\text{IR},t_o},\boldsymbol{\lambda}\right)$ as
\begin{align}
	\phi^\text{IR}\left(\textbf{u}_{\text{IR},t_o},\boldsymbol{\lambda}\right)=
	\frac{\phi_1^\text{IR}\left(\textbf{u}_{\text{IR},t_o}\right) \phi_2^\text{IR}\left(\textbf{u}_{\text{IR},t_o},\boldsymbol{\lambda}\right) \phi_3^\text{IR}\left(\textbf{u}_{\text{IR},t_o},\boldsymbol{\lambda}\right)} {\phi_4^\text{IR}\left(\textbf{u}_{\text{IR},t_o}\right) \phi_5^\text{IR}\left(\textbf{u}_{\text{IR},t_o},\boldsymbol{\lambda}\right) \phi_6^\text{IR}\left(\textbf{u}_{\text{IR},t_o},\boldsymbol{\lambda}\right)}
\end{align}
where 
\begin{align}
	\phi_1^\text{IR}\hspace{-1mm}\left(\textbf{u}_{\text{IR},t_o}\right) &\hspace{-0.7mm}=\hspace{-0.7mm} \prod_{k=1}^{K_t} \hspace{-0.5mm} \textbf{u}_{\text{IR},t_o}^\text{H} \textbf{D}_{k,t_o}\hspace{-0.5mm}(\hspace{-0.2mm}t_o\hspace{-0.3mm}) \textbf{u}_{\text{IR},t_o} , \,\,\,
	\phi_2^\text{IR}\hspace{-1mm}\left(\hspace{-0.2mm}\textbf{u}_{\text{IR},t_o},\hspace{-0.3mm}\boldsymbol{\lambda}\hspace{-0.3mm}\right) \hspace{-0.8mm} = \hspace{-1.2mm} \prod_{k=K_t+1}^{K_t+K_s} \hspace{-1.5mm} \left(\hspace{-0.2mm}\textbf{u}_{\text{IR},t_o}^\text{H} \textbf{D}_{k,t_o}\hspace{-0.5mm}(t_o) \textbf{u}_{\text{IR},t_o}\hspace{-0.3mm}\right)^{\hspace{-0.2mm}\lambda_k\hspace{-0.2mm} \left(\hspace{-0.2mm}1-f_{k,t_o}\hspace{-0.2mm}\left(\tilde{\gamma}_k(t_o)\right)\hspace{-0.3mm}\right) } ,  \nonumber \\
	\phi_3^\text{IR}\hspace{-1mm}\left(\hspace{-0.2mm}\textbf{u}_{\text{IR},t_o},\boldsymbol{\lambda}\hspace{-0.3mm}\right) &\hspace{-0.7mm}=\hspace{-0.7mm} \prod_{k=K_t+1}^{K_t+K_s} \left\{ \prod_{t=t_o+1}^{T} \left(\textbf{u}_{\text{IR},t_o}^\text{H} \tilde{\textbf{D}}_{k,t}(t) \textbf{u}_{\text{IR},t_o}\right)^{1-f_{k,t}\left(\tilde{\gamma}_k(t)\right)} \right\}^{\lambda_k} , \nonumber \\
	\phi_4^\text{IR}\hspace{-1mm}\left(\textbf{u}_{\text{IR},t_o}\right) &\hspace{-0.7mm}=\hspace{-0.7mm} \prod_{k=1}^{K_t} \textbf{u}_{\text{IR},t_o}^\text{H} \textbf{E}_{k,t_o}\hspace{-0.5mm}(t_o) \textbf{u}_{\text{IR},t_o} , \,\,\,
	\phi_5^\text{IR}\hspace{-1mm}\left(\hspace{-0.2mm}\textbf{u}_{\text{IR},t_o},\hspace{-0.3mm}\boldsymbol{\lambda}\hspace{-0.3mm}\right) \hspace{-0.8mm} = \hspace{-1.2mm} \prod_{k=K_t+1}^{K_t+K_s} \hspace{-1.5mm} \left(\hspace{-0.2mm}\textbf{u}_{\text{IR},t_o}^\text{H} \textbf{E}_{k,t_o}\hspace{-0.5mm}(t_o) \textbf{u}_{\text{IR},t_o}\hspace{-0.3mm}\right)^{\hspace{-0.2mm}\lambda_k\hspace{-0.2mm} \left(\hspace{-0.2mm}1-f_{k,t_o}\hspace{-0.2mm}\left(\tilde{\gamma}_k(t_o)\right)\hspace{-0.3mm}\right) } ,  \nonumber \\
	\phi_6^\text{IR}\hspace{-1mm}\left(\hspace{-0.2mm}\textbf{u}_{\text{IR},t_o},\boldsymbol{\lambda}\hspace{-0.3mm}\right) &\hspace{-0.7mm}=\hspace{-0.7mm} \prod_{k=K_t+1}^{K_t+K_s} \left\{ \prod_{t=t_o+1}^{T} \left(\textbf{u}_{\text{IR},t_o}^\text{H} \tilde{\textbf{E}}_{k,t}(t) \textbf{u}_{\text{IR},t_o}\right)^{1-f_{k,t}\left(\tilde{\gamma}_k(t)\right)} \right\}^{\lambda_k} .
\end{align}
\begin{theorem} \label{theorm:first-order_HARQ}
	The first-order optimality condition of the optimization problem in \eqref{eq:problem_HARQ_sec} at the $t_o$-th transmission round ($2 \le t_o <T$) satisfies  when
	\begin{align}
	\bar{\textbf{A}}^\text{IR}\left(\textbf{u}_{\text{IR},t_o},\boldsymbol{\lambda}\right) \textbf{u}_{\text{IR},t_o}
	= \phi^\text{IR}\left(\textbf{u}_{\text{IR},t_o},\boldsymbol{\lambda}\right) \bar{\textbf{B}}^\text{IR}\left(\textbf{u}_{\text{IR},t_o},\boldsymbol{\lambda}\right) \textbf{u}_{\text{IR},t_o}  \label{eq:KKT_first_HARQ}
	\end{align}
	where $\bar{\textbf{A}}^\text{IR}\left(\textbf{u}_{\text{IR},t_o},\boldsymbol{\lambda}\right)$ and $\bar{\textbf{B}}^\text{IR}\left(\textbf{u}_{\text{IR},t_o},\boldsymbol{\lambda}\right)$ are given by

	\begin{align}
	&\bar{\textbf{A}}^\text{IR} \left(\textbf{u}_{\text{IR},t_o},\boldsymbol{\lambda}\right) =
	\prod_{k=1}^{K_t} \textbf{u}_{\text{IR},t_o}^\text{H} \textbf{D}_{k,t_o}(t_o) \textbf{u}_{\text{IR},t_o}
	\prod_{k=K_t+1}^{K_t+K_s} \left(\textbf{u}_{\text{IR},t_o}^\text{H} \textbf{D}_{k,t_o}(t_o) \textbf{u}_{\text{IR},t_o}\right)^{\lambda_k \left(1-f_{k,t_o}\left(\tilde{\gamma}_k(t_o)\right)\right) } \nonumber \\
	& \times
	\prod_{k=K_t+1}^{K_t+K_s} \left\{ \prod_{t=t_o+1}^{T}  \left(\textbf{u}_{\text{IR},t_o}^\text{H} \tilde{\textbf{D}}_{k,t}(t) \textbf{u}_{\text{IR},t_o}\right)^{1-f_{k,t}\left(\tilde{\gamma}_k(t)\right)} \right\}^{\lambda_k} 
	\left\{
	\sum_{k=1}^{K_t} \frac{2 \textbf{D}_{k,t_o}(t_o)} {\textbf{u}_{\text{IR},t_o}^\text{H} \textbf{D}_{k,t_o}(t_o) \textbf{u}_{\text{IR},t_o}}
	\right.
	\nonumber \\
	&\left.
	+ \sum_{k=K_t+1}^{K_t+K_s} 
	\frac{2 \lambda_k \left(1 - f_{k,t_o} \left(\tilde{\gamma}_k(t_o)\right)\right) \textbf{D}_{k,t_o}(t_o)} {\textbf{u}_{\text{IR},t_o}^\text{H} \textbf{D}_{k,t_o}(t_o) \textbf{u}_{\text{IR},t_o}}
	+ \sum_{k=K_t+1}^{K_t+K_s} \sum_{t=t_o+1}^{T} \frac{2 \lambda_k \left(1 - f_{k,t} \left(\tilde{\gamma}_k(t)\right)\right) \tilde{\textbf{D}}_{k,t}(t)} {\textbf{u}_{\text{IR},t_o}^\text{H} \tilde{\textbf{D}}_{k,t}(t) \textbf{u}_{\text{IR},t_o}} 
	\right\}, \label{eq:bar_A_HARQ} \\
	&\bar{\textbf{B}}^\text{IR}  \left(\textbf{u}_{\text{IR},t_o},\boldsymbol{\lambda}\right) = 
	\prod_{k=1}^{K_t} \textbf{u}_{\text{IR},t_o}^\text{H} \textbf{E}_{k,t_o}(t_o) \textbf{u}_{\text{IR},t_o}
	\prod_{k=K_t+1}^{K_t+K_s} \left\{\textbf{u}_{\text{IR},t_o}^\text{H} \textbf{E}_{k,t_o}(t_o) \textbf{u}_{\text{IR},t_o}\right\}^{\lambda_k \left(1-f_{k,t_o}\left(\tilde{\gamma}_k(t_o)\right)\right) } \nonumber \\
	& \times
	\prod_{k=K_t+1}^{K_t+K_s} \left\{ \prod_{t=t_o+1}^{T} \left(\textbf{u}_{\text{IR},t_o}^\text{H} \tilde{\textbf{E}}_{k,t}(t) \textbf{u}_{\text{IR},t_o}\right)^{1-f_{k,t}\left(\tilde{\gamma}_k(t)\right)} \right\}^{\lambda_k}
	\left\{
	\sum_{k=1}^{K_t} \frac{2 \textbf{E}_{k,t_o}(t_o)} {\textbf{u}_{\text{IR},t_o}^\text{H} \textbf{E}_{k,t_o}(t_o) \textbf{u}_{\text{IR},t_o}} \right.
	\nonumber \\
	&\left. +
	\sum_{k=K_t+1}^{K_t+K_s} 
	\frac{2 \lambda_k \left(1 - f_{k,t_o}\left(\tilde{\gamma}_k(t_o)\right)\right) \textbf{E}_{k,t_o}(t_o)} {\textbf{u}_{\text{IR},t_o}^\text{H} \textbf{E}_{k,t_o}(t_o) \textbf{u}_{\text{IR},t_o}} 
	+ \sum_{k=K_t+1}^{K_t+K_s} \sum_{t=t_o+1}^{T} \frac{2 \lambda_k \left(1 - f_{k,t}\left(\tilde{\gamma}_k(t)\right)\right) \tilde{\textbf{E}}_{k,t}(t)} {\textbf{u}_{\text{IR},t_o}^\text{H} \tilde{\textbf{E}}_{k,t}(t) \textbf{u}_{\text{IR},t_o}}
	\right\} . \label{eq:bar_B_HARQ}
	\end{align}
	In addition, the Lagrangian multipliers $\boldsymbol{\lambda}$ are chosen so that $\textbf{u}_{\text{IR},t_o}$ satisfies
	\begin{align}
	\tilde{R}_{k,t_o}\left(\textbf{u}_{\text{IR},t_o}, \tilde{\gamma}_k(t_o)\right) + \sum_{t=t_o+1}^{T} \hat{R}_{k,t}\left(\textbf{u}_{\text{IR},t_o}, \tilde{\gamma}_k(t)\right)
	+ \sum_{t=1}^{t_o-1} \bar{R}_k(t)  = \frac{D_s}{\delta_k} , \quad  k \in \mathcal{K}_s . \label{eq:Lagrangian_IR}
	\end{align}
	If the equation \eqref{eq:Lagrangian_IR} cannot be satisfied, $\boldsymbol{\lambda}$ are chosen so that $\textbf{u}_{\text{IR},t_o}$ satisfies $\tilde{R}_{k,t_o}\left(\textbf{u}_{\text{IR},t_o}, \tilde{\gamma}_k(t_o)\right) + \sum_{t=t_o+1}^{T} \hat{R}_{k,t}\left(\textbf{u}_{\text{IR},t_o}, \tilde{\gamma}_k(t)\right)
	+ \sum_{t=1}^{t_o-1} \bar{R}_k(t)  > \frac{D_s}{\delta_k} , \,  k \in \mathcal{K}_s$.
\end{theorem}
\begin{IEEEproof}
	This can be readily proven using the approach in the proof of Theorem \ref{theorm:first-order}.
\end{IEEEproof}

From Theorem \ref{theorm:first-order_HARQ}, for the first transmission round ($t_o=1$),  we can readily obtain the first-order optimality condition for the precoding vector by using $t_o=1$ in \eqref{eq:KKT_first_HARQ}. 
In addition, Lagrangian multipliers $\boldsymbol{\lambda}$ are chosen so that $\textbf{u}_\text{IR}$ satisfies an equation in \eqref{eq:Lagrangian_IR} with $t_o\hspace{-0.5mm}=\hspace{-0.5mm}1$ and $\sum_{t=1}^{t_o-1} \bar{R}_k(t) \hspace{-0.5mm} = \hspace{-0.5mm} 0$.
If the equation \eqref{eq:Lagrangian_IR} with $t_o=1$ and $\sum_{t=1}^{t_o-1} \bar{R}_k(t) = 0$ cannot be satisfied, $\boldsymbol{\lambda}$ are chosen so that $\textbf{u}_\text{IR}$ satisfies $\tilde{R}_{k,1}\left(\textbf{u}_\text{IR}, \tilde{\gamma}_k(1)\right) + \sum_{t=2}^{T} \hat{R}_{k,t}\left(\textbf{u}_\text{IR}, \tilde{\gamma}_k(t)\right) > \frac{D_s}{\delta_k} , \,  k \in \mathcal{K}_s$.

In addition, at the last transmission round, since there is no future transmission round, the functions related to $\hat{R}_{k,t}\left(\textbf{u}_{\text{IR},t_o}, \tilde{\gamma}_k(t)\right)$ (e.g., $\phi_3^\text{IR}\left(\textbf{u}_{\text{IR},t_o},\boldsymbol{\lambda}\right)$ and $\phi_6^\text{IR}\left(\textbf{u}_{\text{IR},t_o},\boldsymbol{\lambda}\right)$) need to be excluded in \eqref{eq:KKT_first_HARQ}.
Therefore, the first-order optimality condition of the optimization problem in \eqref{eq:problem_HARQ_Last} for the precoding vector is given by
\begin{align}
	\bar{\textbf{A}}^\text{IR} \left(\textbf{u}_{\text{IR},T},\boldsymbol{\lambda}\right) \textbf{u}_{\text{IR},T}
	= \phi^\text{IR}(\textbf{u}_{\text{IR},T},\boldsymbol{\lambda}) \bar{\textbf{B}}^\text{IR} \left(\textbf{u}_{\text{IR},T},\boldsymbol{\lambda}\right) \textbf{u}_{\text{IR},T} \label{eq:KKT_first_HARQ_Last}
\end{align}
where $\phi^\text{IR}(\textbf{u}_{\text{IR},T},\boldsymbol{\lambda})=\frac{\phi_1^\text{IR}(\textbf{u}_{\text{IR},T}) \phi_2^\text{IR}(\textbf{u}_{\text{IR},T},\boldsymbol{\lambda})} {\phi_4^\text{IR}(\textbf{u}_{\text{IR},T}) \phi_5^\text{IR}(\textbf{u}_{\text{IR},T},\boldsymbol{\lambda})}$.
In \eqref{eq:KKT_first_HARQ_Last}, $\bar{\textbf{A}}^\text{IR} \left(\textbf{u}_{\text{IR},T},\boldsymbol{\lambda}\right)$ and $\bar{\textbf{B}}^\text{IR} \left(\textbf{u}_{\text{IR},T},\boldsymbol{\lambda}\right)$ are given by

\begin{align}
	\bar{\textbf{A}}^\text{IR} \left(\textbf{u}_{\text{IR},T},\boldsymbol{\lambda}\right) &=
	\prod_{k=1}^{K_t} \textbf{u}^\text{H}_{\text{IR},T} \textbf{D}_{k,T}\left(T\right) \textbf{u}_{\text{IR},T}
	\prod_{k=K_t+1}^{K_t+K_s} \left(\textbf{u}^\text{H}_{\text{IR},T} \textbf{D}_{k,T}\left(T\right) \textbf{u}_{\text{IR},T}\right)^{\lambda_k \left(1-f_{k,T}\left(\tilde{\gamma}_k\left(T\right)\right)\right) } \nonumber \\
	&\quad \times
	\left\{
	\sum_{k=1}^{K_t} \frac{2 \textbf{D}_{k,T}(T)} {\textbf{u}^\text{H}_{\text{IR},T} \textbf{D}_{k,T}\left(T\right) \textbf{u}_{\text{IR},T}}
	+ \sum_{k=K_t+1}^{K_t+K_s} \frac{2 \lambda_k \left(1-f_{k,T}\left(\tilde{\gamma}_k\left(T\right)\right)\right) \textbf{D}_{k,T}\left(T\right)} {\textbf{u}^\text{H}_{\text{IR},T} \textbf{D}_{k,T}\left(T\right) \textbf{u}_{\text{IR},T}} \right\}, \label{eq:bar_A_HARQ_Last} \\
	\bar{\textbf{B}}^\text{IR} \left(\textbf{u}_{\text{IR},T},\boldsymbol{\lambda}\right) &=
	\prod_{k=1}^{K_t} \textbf{u}^\text{H}_{\text{IR},T} \textbf{E}_{k,T}\left(T\right) \textbf{u}_{\text{IR},T}
	\prod_{k=K_t+1}^{K_t+K_s} \left(\textbf{u}^\text{H}_{\text{IR},T} \textbf{E}_{k,T}\left(T\right) \textbf{u}_{\text{IR},T}\right)^{\lambda_k \left(1-f_{k,T}\left(\tilde{\gamma}_k\left(T\right)\right)\right) } \nonumber \\
	&\quad \times
	\left\{
	\sum_{k=1}^{K_t} \frac{2 \textbf{E}_{k,T}(T)} {\textbf{u}^\text{H}_{\text{IR},T} \textbf{E}_{k,T}\left(T\right) \textbf{u}_{\text{IR},T}}
	+ \sum_{k=K_t+1}^{K_t+K_s} \frac{2 \lambda_k \left(1-f_{k,T}\left(\tilde{\gamma}_k\left(T\right)\right)\right) \textbf{E}_{k,T}\left(T\right)} {\textbf{u}^\text{H}_{\text{IR},T} \textbf{E}_{k,T}\left(T\right) \textbf{u}_{\text{IR},T}} \right\} . \label{eq:bar_B_HARQ_Last}
\end{align}
The Lagrangian multipliers $\boldsymbol{\lambda}$ are chosen so that $\textbf{u}_\text{IR}$ satisfies
\begin{align}
	\tilde{R}_{k,T}\left(\textbf{u}_{\text{IR},T}, \tilde{\gamma}_k(T)\right)
	+ \sum_{t=1}^{T-1} \bar{R}_k(t) = \frac{D_s}{\delta_k} , \quad  k \in \mathcal{K}_s . \label{eq:Lagrangian_last}
\end{align}
If the equation \eqref{eq:Lagrangian_last} cannot be satisfied, $\boldsymbol{\lambda}$ are chosen so that $\textbf{u}_{\text{IR},T}$ satisfies $\tilde{R}_{k,T}\left(\textbf{u}_{\text{IR},T}, \tilde{\gamma}_k(T)\right)
+ \sum_{t=1}^{T-1} \bar{R}_k(t)  > \frac{D_s}{\delta_k} ,  k \in \mathcal{K}_s$.

\begin{algorithm} [t]
	\caption{HARQ-GPI} \label{alg:Algorithm2} 
	\begin{algorithmic} 
		\State {\bf{initialize}}: $\textbf{u}_\text{IR}^{(0)}=\text{RZF}$, $\textbf{u}_\text{IR}^{(-1)}=\textbf{0}$, $\lambda_k^{(0)}$, and $\xi$.
		\State Set the iteration count $n = 0$ and $j = 0$.
		%\State {(\bf{The precoding update phase})}
		\FOR {$t=1$ to $T$}
		\WHILE {$ \tilde{R}_{k,t}\left(\textbf{u}_{\text{IR},t}^{(j)}, \tilde{\gamma}_k(t)\right) 
		+ \sum_{i=t+1}^{T} \hat{R}_{k,i} \left(\textbf{u}_{\text{IR},t}^{(j)}, \tilde{\gamma}_k(i)\right)
		+ \sum_{i=1}^{t-1}\bar{R}_k(i)
		< \frac{D_s}{\delta_k} , \quad \forall k \in \mathcal{K}_s$}
		\State $n \leftarrow n+1$.
		\State $\lambda_k^{(n)} \leftarrow 
		\left[\lambda_k^{(n-1)} + \Delta\lambda_k^{(n)} \right]^+  , \quad \forall k \in \mathcal{K}_s$.
		\WHILE {$\left\|\textbf{u}_{\text{IR},t}^{(j)} - \textbf{u}_{\text{IR},t}^{(j-1)} \right\| > \xi$}
		\State $j \leftarrow j+1$.
		\State Create matrices $\bar{\textbf{A}}^\text{IR} \left(\textbf{u}_{\text{IR},t}^{(j-1)},\boldsymbol{\lambda}^{(n)}\right)$ and $\bar{\textbf{B}}^\text{IR} \left(\textbf{u}_{\text{IR},t}^{(j-1)},\boldsymbol{\lambda}^{(n)}\right)$ by using \eqref{eq:bar_A_HARQ} and \eqref{eq:bar_B_HARQ}. 
		\State Update $\textbf{u}_{\text{IR},t}^{(j)} = \left[\bar{\textbf{B}}^\text{IR} \left(\textbf{u}_{\text{IR},t}^{(j-1)},\boldsymbol{\lambda}^{(n)}\right)\right]^{-1} \bar{\textbf{A}}^\text{IR} \left(\textbf{u}_{\text{IR},t}^{(j-1)},\boldsymbol{\lambda}^{(n)}\right) \textbf{u}_{\text{IR},t}^{(j-1)}$.
		\State Normalize $\textbf{u}_{\text{IR},t}^{(j)} = \textbf{u}_{\text{IR},t}^{(j)}/\left\| \textbf{u}_{\text{IR},t}^{(j)}\right\|$. 
		%\State Update {$R_{{\rm e}, m} $} as in \eqref{eq:wiretap_se} with $\bar {\bf{f}}^{(t)}$ for $m \in \CMcal{M}$
		%\State Update {$w_m $} as in \eqref{eq:weight}
		\ENDWHILE
		\ENDWHILE
		\State Update $\textbf{u}_\text{IR}^{\star}(t) = \textbf{u}_\text{IR}^{(j)}(t)$. 
		\State Store the power ratio of $\left\|\textbf{u}_\text{IR}^{(j)}(t)\right\|$ to $\left\|\textbf{u}_{\text{IR},t}^{(j)}\right\|$ as $q(t)$.
		\State Define $\textbf{u}_\text{IR}^{(j)}(t+1)$ by subtracting $\textbf{u}_\text{IR}^{(j)}(t)$ from $\textbf{u}_{\text{IR},t}^{(j)}$.
		\State Normalize $\textbf{u}_{\text{IR},t+1}^{(j)} = \textbf{u}_{\text{IR},t+1}^{(j)}/\left\| \textbf{u}_{\text{IR},t+1}^{(j)}\right\|$. 
		\ENDFOR
		\State Combine $\textbf{u}_\text{IR}^{\star} = \left[\textbf{u}_\text{IR}^{\star}(1)q(1), \textbf{u}_\text{IR}^{\star}(2)(1-q(1))q(2),  \textbf{u}_\text{IR}^{\star}(3)(1-q(1))(1-q(2))q(3), \cdots \right]^\text{H}$
		\State Normalize $\textbf{u}_\text{IR}^{\star} = \textbf{u}_\text{IR}^{\star}/\left\| \textbf{u}_\text{IR}^{\star}\right\|$.
		%\State {\bf{output}}: $\lfloor \left(b_n^{\star}\right)^+ \rfloor$ for $n \in \{1,...,N\}$
		\State {\bf{output}}: $\textbf{u}_\text{IR}^{\star}$
	\end{algorithmic}
\end{algorithm}

\subsection{Precoding Algorithm}
In this subsection, by modifying the Algorithm \ref{alg:Algorithm1}, we develop an algorithm that finds the principal eigenvector of generalized eigenvalue problems in \eqref{eq:KKT_first_HARQ} and \eqref{eq:KKT_first_HARQ_Last} as presented in Algorithm \ref{alg:Algorithm2}.
We denote a proposed algorithm as the HARQ-GPI.
Compared to the Delay-GPI, we recursively optimize the precoding vector of the \ac{IR-HARQ} scheme.
Specifically, in the $t$-th transmission round, when the precoding vector $\textbf{u}_{\text{IR},t} = \left[\textbf{u}_\text{IR}^\text{H}(t),\cdots,\textbf{u}_\text{IR}^\text{H}(T)\right]^\text{H}$ converges to a principal eigenvector, we only stores the precoding vector of the corresponding transmission round $\textbf{u}_\text{IR}(t) = \left[\textbf{u}_1^\text{H}(t),\cdots,\textbf{u}_{K_t+K_s}^\text{H}(t)\right]^\text{H}$ and the power ratio of $\left\|\textbf{u}_\text{IR}(t)\right\|$ to $\left\|\textbf{u}_{\text{IR},t}\right\|$.
In the next transmission round, by eliminating $\textbf{u}_\text{IR}(t)$, we optimize the new precoding vector $\textbf{u}_{\text{IR},t+1} \hspace{-1mm} = \hspace{-1mm} \left[\textbf{u}_\text{IR}^\text{H}(t\hspace{-0.6mm}+\hspace{-0.8mm}1),\cdots,\textbf{u}_\text{IR}^\text{H}(T)\right]^\text{H}$.
Finally, after optimizing a precoding vector of the last transmission round, we multiply the power ratio occupied in each transmission round by the optimal precoding vector of each transmission round and combine them into $\textbf{u}_\text{IR}^{\star}$, and normalize it.

%
%%---------------------------------------------------------------------------%
%%                          Sec: Numerical Result                            %
%%---------------------------------------------------------------------------%
%
\section{Numerical Results} \label{sec:Numerical}
In this section, we evaluate the sum spectral efficiency according to the network parameters (e.g., user number, blocklength, and transmission round).
We use the one-ring model for the spatial covariance matrix of the channel \cite{Cle:13}.
Specifically, we consider that the \ac{BS} is equipped with uniform circular array with a circle of radius $\nu D$ where $\nu$ is the wavelength and $D=\frac{0.5}{\sqrt{\left(1-\cos(2\pi/N)\right)^2 + \sin(2\pi/N)^2}}$.
From this, the channel correlation coefficient between the $n$-th antenna and the $m$-th antenna of the user $k$ is given by
\begin{align}
\left[\textbf{C}_k\right]_{n,m} = \frac{1}{2\Delta} \int_{\theta_k - \Delta}^{\theta_k + \Delta} \exp\left\{-j \frac{2 \pi}{\nu} \Psi(x) (\textbf{r}_n - \textbf{r}_m)\right\} \, dx
\end{align}
where $\theta_k$ is the \ac{AoA} of the user $k$, $\Delta$ is the angular spread, $\Psi(x)=[\cos(x), \sin(x)]$ is the wave vector for a planer wave impinging with the angle of $x$, and $\textbf{r}_n$ is the position vector for the $n$-th antenna of the \ac{BS}.
We then assume that the \ac{AoA} of the user $k$ is determined by its spatial location, which is uniformly distributed in $(0,2\pi]$.
\begin{table}[!t]
	\caption{Parameter values if not otherwise specified \label{table:parameter}} 
	\begin{center}
		\rowcolors{2}%{green!20!yellow}
		{cyan!15!}{}
		\renewcommand{\arraystretch}{1.5}
		\begin{tabular}{l l | l l}
			\hline 
			{\bf Parameters} & {\bf Values} & {\bf Parameters} & {\hspace{0.32cm}}{\bf Values} \\
			\hline 
			\hspace{0.15cm}$\sigma^2$ [dB]  & \hspace{0.2cm}$-113$ 
			& \hspace{0.12cm}$\Delta$ & \hspace{0.2cm}$\frac{\pi}{6}$ \\ %\addlinespace
			\hspace{0.15cm}$m$ & \hspace{0.2cm}$100$ 
			& \hspace{0.12cm}$\varepsilon$ & \hspace{0.2cm}$10^{-5}$  \\ %\addlinespace
			\hspace{0.2cm}$\xi$  & \hspace{0.2cm}$0.05$ 
			& \hspace{0.12cm}$N$   & \hspace{0.2cm}$8$  \\ %\addlinespace
			\hspace{0.2cm}$K_t$  & \hspace{0.2cm}$3$ 
			& \hspace{0.12cm}$D_s$ [byte]   & \hspace{0.2cm}$32$  \\ %\addlinespace
			\hline
		\end{tabular}
	\end{center}
\end{table}%
In addition, for a given channel realization, since the algorithm can fail, we define the sum spectral efficiency as
\begin{align}
	R_s = 
	\left\{
	\begin{aligned}
	 &\,\, \sum_{k=1}^{K_t} w_k R_k(\textbf{u}) +  \sum_{k=K_t+1}^{K_t+K_s} \frac{w_k D_s} {\delta_k} , \quad \text{if \textbf{u} is feasible} \nonumber \\
	 &\,\, 0 ,     \quad \text{otherwise}
	 \end{aligned}
	 \right.
\end{align}
where $w_k$ is the weight allocated to a user $k$.
If the algorithm fails or the precoding vector is infeasible, we set the sum spectral efficiency for a given channel realization to zero.
Furthermore, even though the algorithm performs well, since baseline schemes (e.g., \ac{MRT} and \ac{RZF} methods) may not satisfy the latency requirement, we modify the sum spectral efficiency as
\begin{align}
R_s = \sum_{k=1}^{K_t} w_k R_k(\textbf{u}) + \sum_{k=K_t + 1}^{K_t + K_s} p_{c,k} \frac{w_k D_s} {\delta_k} \label{eq:sum_base}
\end{align}
where $p_{c,k}$ is the binary variable, which indicates that the latency requirement of user $k$ is satisfied if $p_{c,k}=1$; otherwise, $p_{c,k}=0$.
We denote the ergodic sum spectral efficiency as $\bar{R}_s = \mathbb{E}\left[R_s\right]$.
Unless otherwise specified, values of simulation parameters presented in Table \ref{table:parameter} are used.

In this simulation, we compare the proposed algorithm with following methods:
\begin{itemize}
	%
%	\item Baseline 1: this method considers that there only exist delay-tolerant users.
%	This algorithm is designed to solve the following problem:
%	%
%	%
%	%
%	\begin{align}
%	%
%	&\underset{\textbf{u}}{\text{maximize}} \quad \sum_{k=1}^{K_{t1}} R_k(\textbf{u})  \nonumber \\
%	%
%	&\text{subject to} \quad \sum_{i=1}^{K_t} \left\lVert \textbf{u}_i \right\rVert \le 1 \nonumber \\
%	%
%	& \quad \quad \quad \quad \quad  
%	\frac{D_s}{R_k(\textbf{u})} < \delta_k , \quad  k = K_{t1}+1,\cdots,K_{t2} \label{eq:baseline1}
%	%
%	\end{align}
%	%
%	%
%	%
%	where $K_t=K_{t1}+K_{t2}$ and $R_k(\textbf{u})
%	= \log_2  \left(\frac{\textbf{u}^\text{H} \textbf{A}_k \textbf{u}} {\textbf{u}^\text{H} \textbf{B}_k \textbf{u}}\right)$.
%	On the delay constriant in \eqref{eq:baseline1}, as the spectral efficiency of the delay-tolerant user is considered, the first-order optimality condition of the optimizaiton problem is newly obtained by reflecting $m=\infty$.
	%
	\item Infinite-GPI: when we solve the optimization problem in \eqref{eq:problem_re},
	we obtain the optimal precoding vector by using the first-order optimality conditions in \eqref{eq:KKT_first} and \eqref{eq:lagrange} with $m=\infty$.	
	However, the spectral efficiency of delay-constrained users with finite blocklength $m$ is still used when checking latency requirements of delay-constrained users.
	\item MRT: this method is a \ac{MRT} precoding \cite{Lo:99}. The precoding vector of user $k$ is the same as $\textbf{u}_k=\textbf{h}_k$.
	In this method, we assume $\tilde{\gamma}_k = \gamma_k$.
	\item RZF: this method is a \ac{RZF} precoding \cite{Wag:12}.
	The precoding vector of user $k$ is given by
	\begin{align}
	\textbf{u}_k = \left(\textbf{H}_c \textbf{H}_c^\text{H} + \frac{\sigma^2}{P} \textbf{I}_N\right)^{-1} \textbf{H}_c
	\end{align}
	where $\textbf{H}_c = [\textbf{h}_1,\cdots,\textbf{h}_{K_t},\textbf{h}_{K_t+1},\cdots,\textbf{h}_{K_t+K_s}] \in \mathbb{C}^{N(K_t+K_s) \times 1}$.
	Suppose that $\tilde{\gamma}_k = \gamma_k$.
\end{itemize}

\subsection{Performance analysis of DCTU-MIMO networks without IR-HARQ}
In this subsection, we analyze the ergodic sum spectral efficiency depending on the network parameters.
%We first explain the latency requirement unit by providing an example.
%When the latency requirement of user $k$ is $\delta_k=500$, the actual latency requirement is obtained by dividing $\delta_k$ into the bandwidth.
%Therefore, if the bandwidth is 5MHz, the actual latency requirement is 0.1ms.
Note that the latency requirement $\delta_k$ is the requirement for the spectral efficiency, not the actual data rate, obtained from the spectral efficiency by multiplying with the bandwidth.
Hence, the actual latency requirement is $\delta_k$ divided by the bandwidth.
For example, $\delta_k=500$ means the actual latency requirement is 0.1ms when the bandwidth is 5MHz.

%
%\/\/\/\/\/\/\/\/\/\/\/\/\/\/\/\/\/\/\/\/\/\/\/\/\/\/\/\/\/\/\/\/\/\/\/\/\/\/\/\/\/\/\/\/\/\/\/\/\/\/\/\/\/\/\/\/\/\/\/\/\/\/\/\/\/\/\/\/\/\/\/\/\/\/\/\/\/\/\/\/\/\/\/\/\/\/\/
%***** x-axis:  [tc][bc][0.7] y-axis: [bc][tc][0.7], legend: [Bl][Bl][0.59]
\begin{figure}[t!]
	\begin{center}   
		{ 
			
			\psfrag{X1}[Bl][Bl][0.85]{$\frac{P}{\sigma^2} \: [\text{dB}]$}
			\psfrag{Y1}[Bl][Bl][0.85]{$\text{Ergodic Sum Spectral Efficiency}, \: \bar{R}_s \: [\text{bits/sec/Hz}]$}
			\psfrag{AAAAAAAAAA1}[Bl][Bl][0.75]{$\text{Delay-GPI}$}
			\psfrag{A2}[Bl][Bl][0.75]{$\text{Infinte-GPI}$}
			\psfrag{A3}[Bl][Bl][0.75]{$\text{RZF}$}
			\psfrag{A4}[Bl][Bl][0.75]{$\text{MRT}$}
			\psfrag{A5}[Bl][Bl][0.75]{$N=6, m=100$}
			\psfrag{A6}[Bl][Bl][0.75]{$N=8, m=100$}
			\psfrag{A7}[Bl][Bl][0.75]{$N=8, m=300$}
			\includegraphics[width=0.6\columnwidth]{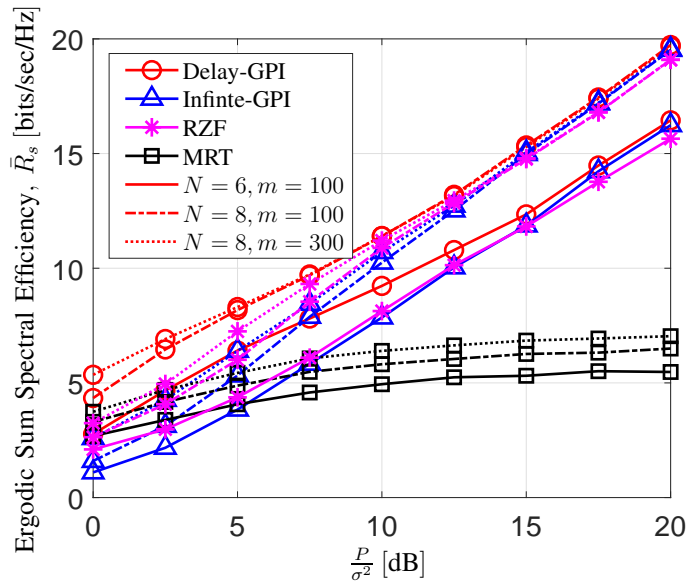}
			\vspace{-10mm}
		}
	\end{center}
	\caption{
		Ergodic sum spectral efficiency $\bar{R}_s$ as a function of $P/\sigma^2$ with $K_t=3$, $K_s=2$, $\delta_4=250$, and $\delta_5=450$ for different values of $N$ and $m$.
		\vspace{-5mm}
	}
	\label{fig:original_antenna_block}
\end{figure}
%\/\/\/\/\/\/\/\/\/\/\/\/\/\/\/\/\/\/\/\/\/\/\/\/\/\/\/\/\/\/\/\/\/\/\/\/\/\/\/\/\/\/\/\/\/\/\/\/\/\/\/\/\/\/\/\/\/\/\/\/\/\/\/\/\/\/\/\/\/\/\/\/\/\/\/\/\/\/\/\/\/\/\/\/\/\/\/
%

Figure~\ref{fig:original_antenna_block} presents the ergodic sum spectral efficiency $\bar{R}_s$ as a function of $P/\sigma^2$ with $K_t=3$, $K_s=2$, $\delta_4=250$, and $\delta_5=450$ for different values of the number of the \ac{BS} antennas $N$ and the blocklength $m$.
Here, $\tilde{\gamma}_4=2.38$, $\tilde{\gamma}_5=1.35$, $w_1=w_2=w_3=1$, and $w_4=w_5=3$.
From this figure, we can know that the Delay-GPI outperforms \ac{MRT} and \ac{RZF} methods.
%This is because, as the principal eigenvector of the baseline 1 is obtained by the first-order optimality condition with $m=\infty$, the obtained principal eigenvector is not optimal.
%Furthermore, the baseline 2 may sometimes be better than the proposed algorithm, but the baseline algorithm does not always satisfy the delay constraint of the delay-constrained user.
%In this simulation, the delay constrained user of the baseline 2 satisfies the delay constraint with about $65\%$.
This is because, since two baseline schemes cannot satisfy the latency requirement of delay-constrained users, the spectral efficiencies of delay-constrained users can be excluded.
In addition, the Delay-GPI is much better than the Infinite-GPI at low $P/\sigma^2$.
This is because, as the principal eigenvector of the Infinite-GPI is obtained by the first-order optimality condition with $m=\infty$, the obtained principal eigenvector is not optimal and can be infeasible.
However, at high $P/\sigma^2$, the Delay-GPI is slightly better than the Infinite-GPI because the probability of success of both algorithms is similar.
We can also see that as the number of the \ac{BS} antennas $N$ increases, the ergodic sum spectral efficiency increases due to larger diversity gain.
Furthermore, we can know that as the blocklength $m$ increases, the ergodic sum spectral efficiency increases.
This is because as $m$ increases, the effect of the channel dispersion decreases and the spectral efficiency of delay-constrained users increases.
Hence, since the spectral efficiency of delay-constrained users increases with larger blocklength, the ergodic sum spectral efficiency increases by allocating more transmission power to the user that has well-conditioned channel gain.
\begin{figure}[t!]
	\begin{center}   
		{ 
			
			\psfrag{X1}[Bl][Bl][0.85]{$\frac{P}{\sigma^2} \: [\text{dB}]$}
			\psfrag{Y1}[Bl][Bl][0.85]{$\text{Ergodic Sum Spectral Efficiency}, \: \bar{R}_s \: [\text{bits/sec/Hz}]$}
			\psfrag{AAAAAAA1}[Bl][Bl][0.75]{$\text{Delay-GPI}$}
			\psfrag{A2}[Bl][Bl][0.75]{$\text{Infinte-GPI}$}
			\psfrag{A3}[Bl][Bl][0.75]{$\text{RZF}$}
			\psfrag{A4}[Bl][Bl][0.75]{$\text{MRT}$}
			\psfrag{A5}[Bl][Bl][0.75]{$K_s=1$}
			\psfrag{A6}[Bl][Bl][0.75]{$K_s=2$}
			\includegraphics[width=0.6\columnwidth]{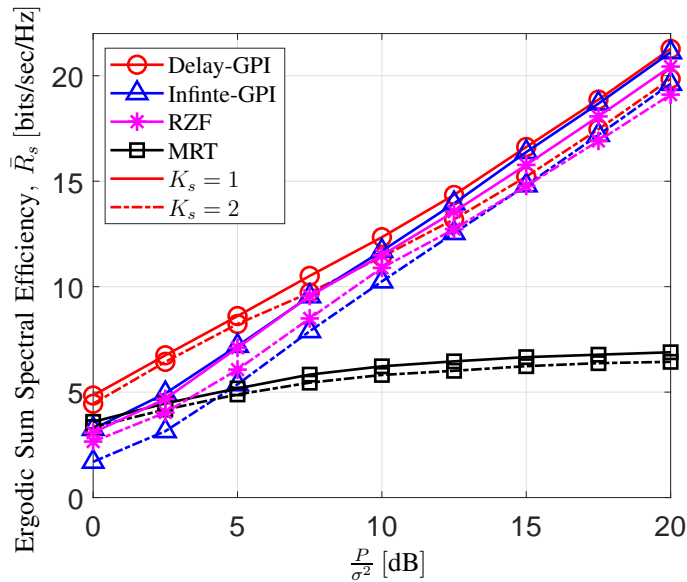}
			\vspace{-10mm}
		}
	\end{center}
	\caption{
		Ergodic sum spectral efficiency $\bar{R}_s$ as a function of $P/\sigma^2$ with $N=8$ and $K_t=3$ for different values of $K_s$.
		In the case of $K_s=1$, $\delta_4=250$.
		When $K_s=2$, $\delta_4=250$ and $\delta_5=450$.
		\vspace{-5mm}
	}
	\label{fig:orginal_user}
\end{figure}
%\/\/\/\/\/\/\/\/\/\/\/\/\/\/\/\/\/\/\/\/\/\/\/\/\/\/\/\/\/\/\/\/\/\/\/\/\/\/\/\/\/\/\/\/\/\/\/\/\/\/\/\/\/\/\/\/\/\/\/\/\/\/\/\/\/\/\/\/\/\/\/\/\/\/\/\/\/\/\/\/\/\/\/\/\/\/\/
%

Figure~\ref{fig:orginal_user} presents the ergodic sum spectral efficiency $\bar{R}_s$ as a function of  $P/\sigma^2$ with $N=8$ and $K_t\hspace{-0.6mm}=\hspace{-0.6mm}3$ for different values of the number of delay-constrained users $K_s$.
Here, when $K_s\hspace{-0.6mm}=\hspace{-0.6mm}1$, $\delta_4\hspace{-0.6mm}=\hspace{-0.6mm}250$, $\tilde{\gamma}_4\hspace{-0.6mm}=\hspace{-0.6mm}2.38$, $w_1\hspace{-0.6mm}=\hspace{-0.6mm}w_2\hspace{-0.6mm}=\hspace{-0.6mm}w_3\hspace{-0.6mm}=\hspace{-0.6mm}1$, and $w_4\hspace{-0.6mm}=\hspace{-0.6mm}3$.
On one hand, when $K_s\hspace{-0.6mm}=\hspace{-0.6mm}2$, $\delta_4\hspace{-0.6mm}=\hspace{-0.6mm}250$, $\delta_5\hspace{-0.6mm}=\hspace{-0.6mm}450$, $\tilde{\gamma}_4\hspace{-0.6mm}=\hspace{-0.6mm}2.38$, $\tilde{\gamma}_5\hspace{-0.6mm}=\hspace{-0.6mm}1.35$, $w_1\hspace{-0.6mm}=\hspace{-0.6mm}w_2\hspace{-0.6mm}=\hspace{-0.6mm}w_3\hspace{-0.6mm}=\hspace{-0.6mm}1$, and $w_4\hspace{-0.6mm}=\hspace{-0.6mm}w_5\hspace{-0.6mm}=\hspace{-0.6mm}3$.
From this figure, we can see that as the number of delay-constrained users $K_s$ increases, the ergodic sum spectral efficiency decreases.
This is because as $K_s$ increases, within limited total transmission power, more transmission power is allocated to delay-constrained users that may have ill-conditioned channel gain to satisfy the latency requirement of delay-constrained users.
Therefore, more transmission power cannot be allocated to the user that has well-conditioned channel gain.
\begin{figure}[t!]
	\begin{center}   
		{ 
			
			\psfrag{X1}[Bl][Bl][0.85]{$\frac{P}{\sigma^2} \: [\text{dB}]$}
			\psfrag{Y1}[Bl][Bl][0.85]{$\text{Ergodic Sum Spectral Efficiency}, \: \bar{R}_s \: [\text{bits/sec/Hz}]$}
			\psfrag{AAAAAAAAA1}[Bl][Bl][0.75]{$\text{Delay-GPI}$}
			\psfrag{A2}[Bl][Bl][0.75]{$\text{Infinte-GPI}$}
			\psfrag{A3}[Bl][Bl][0.75]{$\text{RZF}$}
			\psfrag{A4}[Bl][Bl][0.75]{$\text{MRT}$}
			\psfrag{A5}[Bl][Bl][0.75]{$w_4=1, w_5=1$}
			\psfrag{A6}[Bl][Bl][0.75]{$w_4=3, w_5=3$}
			\includegraphics[width=0.6\columnwidth]{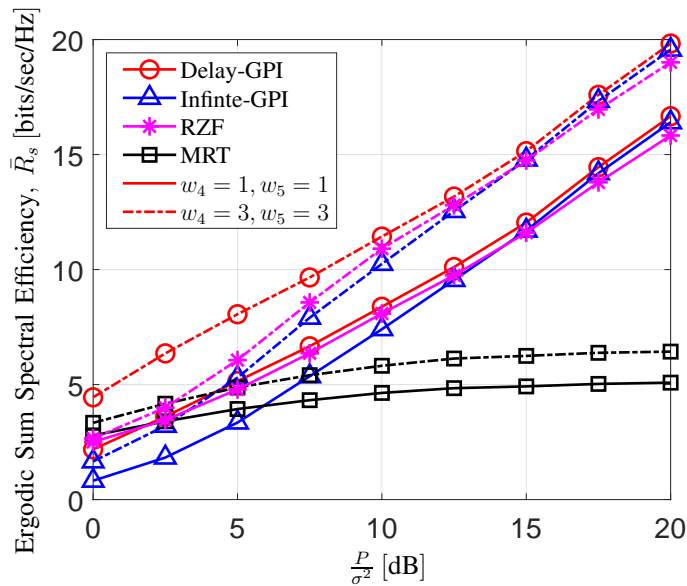}
			\vspace{-10mm}
		}
	\end{center}
	\caption{
		Ergodic sum spectral efficiency $\bar{R}_s$ as a function of $P/\sigma^2$ with $N=8$, $K_t=3$, $K_s=2$, $\delta_4=250$, $\delta_5=450$, $\tilde{\gamma}_4=2.38$, and $\tilde{\gamma}_5=1.35$ for different values of $w_k$.
		\vspace{-5mm}
	}
	\label{fig:original_weight}
\end{figure}
%\/\/\/\/\/\/\/\/\/\/\/\/\/\/\/\/\/\/\/\/\/\/\/\/\/\/\/\/\/\/\/\/\/\/\/\/\/\/\/\/\/\/\/\/\/\/\/\/\/\/\/\/\/\/\/\/\/\/\/\/\/\/\/\/\/\/\/\/\/\/\/\/\/\/\/\/\/\/\/\/\/\/\/\/\/\/\/
%

Figure~\ref{fig:original_weight} presents the ergodic sum spectral efficiency $\bar{R}_s$ as a function of $P/\sigma^2$ with $N=8$, $K_t=3$, $K_s=2$, $w_1=w_2=w_3=1$, $\delta_4=250$, $\delta_5=450$, $\tilde{\gamma}_4=2.38$, and $\tilde{\gamma}_5=1.35$ for different values of the weight allocated to delay-constrained users, i.e., $w_4$ and $w_5$.
From this figure, we can know that when $w_4=w_5=3$, the Delay-GPI is better than other methods in general power regime.
We can also see that as $w_4$ and $w_5$ decreases (e.g., from $w_4=w_5=3$ to $w_4=w_5=1$), even though the difference between the Delay-GPI and other methods decreases, the Delay-GPI is better than other methods in most power regime.
However, when $P/\sigma^2$ is small (e.g., $P/\sigma^2 < 2.5$dB), the \ac{MRT} and \ac{RZF} methods are slightly better than the Delay-GPI.
This is because the definition of the sum spectral efficiency is different.
Specifically, if the Delay-GPI fails, the sum spectral efficiency will be zero.
On the other hand, the \ac{MRT} and \ac{RZF} methods achieve the sum spectral efficiency in \eqref{eq:sum_base} regardless of the algorithm failure (i.e., even when the latency constraint is not satisfied).
For example, when $P/\sigma^2=0$dB, the sum spectral efficiency of the Delay-GPI becomes zero with the algorithm failure probability of 0.3.
On the other hand, for the \ac{MRT} and \ac{RZF} methods, the sum of spectral efficiencies of delay-tolerant users is always achieved even if the delay-constrained users do not satisfy the latency requirement with the probabilities $p_{c,4}=0.1$ and $p_{c,5}=0.26$ in the \ac{MRT} case.
From these result, we can know that when a low power regime, even though the ergodic sum spectral efficiency of the Delay-GPI can be lower than that of the \ac{MRT} and \ac{RZF} methods, the Delay-GPI has the advantage of serving all users because it satisfies the latency constraint of the delay-constrained users that \ac{MRT} and \ac{RZF} methods may not satisfy.

\subsection{Performance analysis of DCTU-MIMO networks with IR-HARQ}
In this subsection, we analyze the effect of transmission rounds and angular spread on the ergodic sum spectral efficiency.
We first introduce methods to compare with the HARQ-GPI.
\begin{itemize}
\item Perfect-GPI: the Perfect-GPI is assumed that a \ac{BS} knows the perfect \ac{CSI} of total transmission rounds in advance. This algorithm is designed to solve the following optimization:
\begin{align}
	&\underset{\textbf{u}_\text{IR}}{\text{maximize}} \quad 
	\sum_{t=1}^{T} \sum_{k=1}^{K_t} R_{k,t}(\textbf{u}_\text{IR}) \nonumber \\
	& \text{subject to} \quad  
	\sum_{t=1}^{T} R_{k,t}(\textbf{u}_\text{IR})  
	\ge \frac{D_s}{\delta_k} , \quad  k \in \mathcal{K}_s . \label{eq:perfect}
\end{align}
%
%
%
%
%Therefore, by supplementing the transmission round to the algorithm \ref{alg:Algorithm1}, we can readily solve above optimization.
The first-order optimality condition of the problem in \eqref{eq:perfect} satisfies when
\begin{align}
	\bar{\textbf{A}}^\text{IR}\left(\textbf{u}_{\text{IR}},\boldsymbol{\lambda}\right) \textbf{u}_{\text{IR}}
	= \phi^\text{IR}\left(\textbf{u}_{\text{IR}},\boldsymbol{\lambda}\right) \bar{\textbf{B}}^\text{IR}\left(\textbf{u}_{\text{IR}},\boldsymbol{\lambda}\right) \textbf{u}_{\text{IR}} . \label{eq:KKT_first_perfect}
\end{align}
In \eqref{eq:KKT_first_perfect}, $\bar{\textbf{A}}^\text{IR}\left(\textbf{u}_{\text{IR}},\boldsymbol{\lambda}\right)$ and $\bar{\textbf{B}}^\text{IR}\left(\textbf{u}_{\text{IR}},\boldsymbol{\lambda}\right)$ are given by
\begin{align}
	\bar{\textbf{A}}^\text{IR} \left(\textbf{u}_{\text{IR}},\boldsymbol{\lambda}\right) &=
	\prod_{t=1}^{T} \prod_{k=1}^{K_t} \textbf{u}_{\text{IR}}^\text{H} \textbf{D}_k(t) \textbf{u}_{\text{IR}}
	\prod_{k=K_t+1}^{K_t+K_s} \left\{ \prod_{t=1}^{T}  \left(\textbf{u}_{\text{IR}}^\text{H} \textbf{D}_k(t) \textbf{u}_{\text{IR}}\right)^{1-f_{k,t}\left(\tilde{\gamma}_k(t)\right)} \right\}^{\lambda_k} 
	\nonumber \\
	& \quad \times
	\left\{
	\sum_{k=1}^{K_t} \sum_{t=1}^{T} \frac{2 \textbf{D}_k(t)} {\textbf{u}_{\text{IR}}^\text{H} \textbf{D}_k(t) \textbf{u}_{\text{IR}}}
	+ \sum_{k=K_t+1}^{K_t+K_s} \sum_{t=1}^{T} \frac{2 \lambda_k \left(1 - f_{k,t} \left(\tilde{\gamma}_k(t)\right)\right) \textbf{D}_k(t)} {\textbf{u}_{\text{IR}}^\text{H} \textbf{D}_k(t) \textbf{u}_{\text{IR}}} 
	\right\}, \label{eq:bar_A_HARQ_perfect} \\
	\bar{\textbf{B}}^\text{IR} \left(\textbf{u}_{\text{IR}},\boldsymbol{\lambda}\right) &=
	\prod_{t=1}^{T} \prod_{k=1}^{K_t} \textbf{u}_{\text{IR}}^\text{H} \textbf{E}_k(t) \textbf{u}_{\text{IR}}
	\prod_{k=K_t+1}^{K_t+K_s} \left\{ \prod_{t=1}^{T}  \left(\textbf{u}_{\text{IR}}^\text{H} \textbf{E}_k(t) \textbf{u}_{\text{IR}}\right)^{1-f_{k,t}\left(\tilde{\gamma}_k(t)\right)} \right\}^{\lambda_k} 
	\nonumber \\
	& \quad \times
	\left\{
	\sum_{k=1}^{K_t} \sum_{t=1}^{T} \frac{2 \textbf{E}_k(t)} {\textbf{u}_{\text{IR}}^\text{H} \textbf{E}_k(t) \textbf{u}_{\text{IR}}}
	+ \sum_{k=K_t+1}^{K_t+K_s} \sum_{t=1}^{T} \frac{2 \lambda_k \left(1 - f_{k,t} \left(\tilde{\gamma}_k(t)\right)\right) \textbf{E}_k(t)} {\textbf{u}_{\text{IR}}^\text{H} \textbf{E}_k(t) \textbf{u}_{\text{IR}}} 
	\right\} . \label{eq:bar_B_HARQ_perfect}
\end{align}
In addition, Lagrangian multipliers $\boldsymbol{\lambda}$ are chosen so that $\textbf{u}_{\text{IR}}$ satisfies
$\sum_{t=1}^{T}\hspace{-0.6mm}\tilde{R}_{k,t}(\textbf{u}_\text{IR}, \tilde{\gamma}_k(t)) \hspace{-0.5mm} \ge \hspace{-0.5mm} \frac{D_s}{\delta_k} , \,  k \in \mathcal{K}_s$.
The algorithm procedure is the same as Algorithm~\ref{alg:Algorithm1}.
\item \ac{MRT}/\ac{RZF}: we assume a \ac{BS} only knows the perfect \ac{CSI} of the current transmission round.
Hence, we perform a \ac{MRT}/\ac{RZF} precoding separately for each transmission round and obtain spectral efficiencies of each transmission round.
After the last transmission round, we sum spectral efficiencies of delay-tolerant users and delay-constrained users for total transmission rounds, respectively, and check the latency requirement for delay-constrained users.
%
%Finally, we add the spectral efficiency of delay-tolerant users and delay-constrained users for total transmission rounds, respectively, and check the latency requirement of delay-constrained users.
%
\end{itemize}
%
%
%
%As mentioned in Section \ref{subsec:SE_B}, unlike the Perfect-GPI, the HARQ-GPI is assumed that the \ac{BS} only knows the perfect \ac{CSI} of the current transmission round and the channel covariance matrix of the future transmission round.
Furthermore, to make a fairness comparison of \ac{MRT} and \ac{RZF} methods with Perfect-GPI and HARQ-GPI, the transmit power at each transmission round of \ac{MRT} and \ac{RZF} methods is $P/T^2$, obtained by dividing the total transmit power $P$ by the square of total transmission rounds $T^2$.

%
%\/\/\/\/\/\/\/\/\/\/\/\/\/\/\/\/\/\/\/\/\/\/\/\/\/\/\/\/\/\/\/\/\/\/\/\/\/\/\/\/\/\/\/\/\/\/\/\/\/\/\/\/\/\/\/\/\/\/\/\/\/\/\/\/\/\/\/\/\/\/\/\/\/\/\/\/\/\/\/\/\/\/\/\/\/\/\/
%***** x-axis:  [tc][bc][0.7] y-axis: [bc][tc][0.7], legend: [Bl][Bl][0.59]
\begin{figure}[t!]
	\begin{center}   
		{ 
			
			\psfrag{X1}[Bl][Bl][0.85]{$\frac{P}{\sigma^2} \: [\text{dB}]$}
			\psfrag{Y1}[Bl][Bl][0.85]{$\text{Ergodic Sum Spectral Efficiency}, \: \bar{R}_s \: [\text{bits/sec/Hz}]$}
			\psfrag{AAAAAAAA1}[Bl][Bl][0.75]{$\text{Perfect-GPI}$}
			\psfrag{A2}[Bl][Bl][0.75]{$\text{HARQ-GPI}$}
			\psfrag{A3}[Bl][Bl][0.75]{$\text{RZF}$}
			\psfrag{A4}[Bl][Bl][0.75]{$\text{MRT}$}
			\psfrag{A5}[Bl][Bl][0.75]{$N=8, T=2$}
			\psfrag{A6}[Bl][Bl][0.75]{$N=8, T=3$}
			\psfrag{A7}[Bl][Bl][0.75]{$N=6, T=3$}
			\includegraphics[width=0.6\columnwidth]{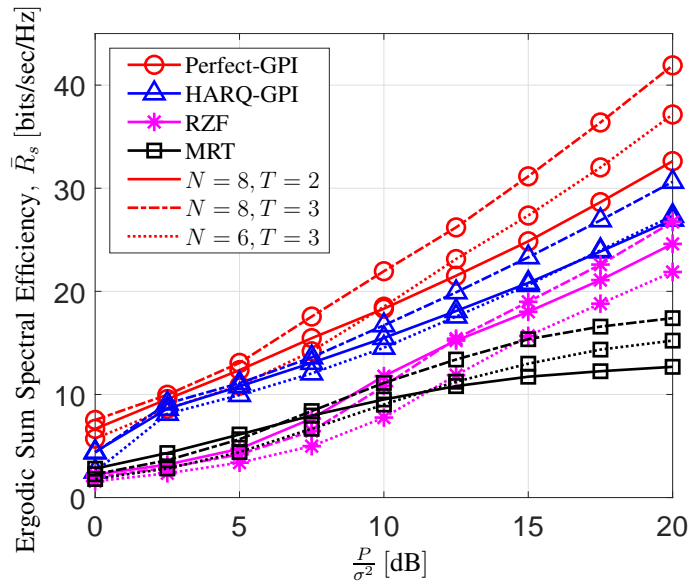}
			\vspace{-10mm}
		}
	\end{center}
	\caption{
		Ergodic sum spectral efficiency $\bar{R}_s$ as a function of $P/\sigma^2$ with $K_t=3$, $K_s=2$, $\delta_4=210$, and $\delta_5=320$ for different values of $N$ and $T$.
	\vspace{-5mm}
	}
	\label{fig:HARQ_antenna_time}
\end{figure}
%\/\/\/\/\/\/\/\/\/\/\/\/\/\/\/\/\/\/\/\/\/\/\/\/\/\/\/\/\/\/\/\/\/\/\/\/\/\/\/\/\/\/\/\/\/\/\/\/\/\/\/\/\/\/\/\/\/\/\/\/\/\/\/\/\/\/\/\/\/\/\/\/\/\/\/\/\/\/\/\/\/\/\/\/\/\/\/
%

Figure~\ref{fig:HARQ_antenna_time} presents the ergodic sum spectral efficiency $\bar{R}_s$ as a function of $P/\sigma^2$ with $K_t=3$, $K_s=2$, $\delta_4=210$, and $\delta_5=320$ for different values of the number of \ac{BS} antennas $N$ and the total transmission rounds $T$.
Here, we use $w_1=w_2=w_3=1$, and $w_4=w_5=3$.
In addition, in the case of $T=2$, $\tilde{\gamma}_4(1)=\tilde{\gamma}_4(2)=1.7$ and $\tilde{\gamma}_5(1)=\tilde{\gamma}_5(2)=1.15$, while when $T=3$, $\tilde{\gamma}_4(1)=\tilde{\gamma}_4(2)=\tilde{\gamma}_4(3)=1.45$ and $\tilde{\gamma}_5(1)=\tilde{\gamma}_5(2)=\tilde{\gamma}_5(3)=1.05$.
From this figure, we can know that the Perfect-GPI outperforms the HARQ-GPI.
This is because the Perfect-GPI knows the perfect \ac{CSI} of total transmission rounds, while the HARQ-GPI knows the prefect \ac{CSI} of the current transmission round and the channel covariance matrix of the future transmission round.
However, the HARQ-GPI is greater than \ac{MRT} and \ac{RZF} methods because the two baseline methods do not have the channel knowledge of future transmission rounds.
We can also see that as the total transmission rounds $T$ increase, the ergodic sum spectral efficiency of the HARQ-GPI increases because of the additional time diversity gain.
However, the ergodic sum spectral efficiency of \ac{MRT} and \ac{RZF} methods decreases with $T$ at small $P/\sigma^2$.
This is because, as $T$ increases, the transmit power at each transmission round decreases, therefore, at small $P/\sigma^2$, even though the opportunity to transmit increases, the latency requirements of delay-constrained users do not be satisfied.
Furthermore, we can know that as the number of \ac{BS} antennas increases, the ergodic sum spectral efficiency increases due to the additional antenna diversity gain.

%
%\/\/\/\/\/\/\/\/\/\/\/\/\/\/\/\/\/\/\/\/\/\/\/\/\/\/\/\/\/\/\/\/\/\/\/\/\/\/\/\/\/\/\/\/\/\/\/\/\/\/\/\/\/\/\/\/\/\/\/\/\/\/\/\/\/\/\/\/\/\/\/\/\/\/\/\/\/\/\/\/\/\/\/\/\/\/\/
%***** x-axis:  [tc][bc][0.7] y-axis: [bc][tc][0.7], legend: [Bl][Bl][0.59]
\begin{figure}[t!]
	\begin{center}   
		{ 
			
			\psfrag{X1}[Bl][Bl][0.85]{$\frac{P}{\sigma^2} \: [\text{dB}]$}
			\psfrag{Y1}[Bl][Bl][0.85]{$\text{Ergodic Sum Spectral Efficiency}, \: \bar{R}_s \: [\text{bits/sec/Hz}]$}
			\psfrag{AAAAAAA1}[Bl][Bl][0.75]{$\text{Perfect-GPI}$}
			\psfrag{A2}[Bl][Bl][0.75]{$\text{HARQ-GPI}$}
			\psfrag{A3}[Bl][Bl][0.75]{$\text{RZF}$}
			\psfrag{A4}[Bl][Bl][0.75]{$\text{MRT}$}
			\psfrag{A5}[Bl][Bl][0.75]{$\Delta=\frac{\pi}{12}$}
			\psfrag{A6}[Bl][Bl][0.75]{$\Delta=\frac{\pi}{6}$}
			\psfrag{A7}[Bl][Bl][0.75]{$\Delta=\frac{\pi}{3}$}
			\includegraphics[width=0.6\columnwidth]{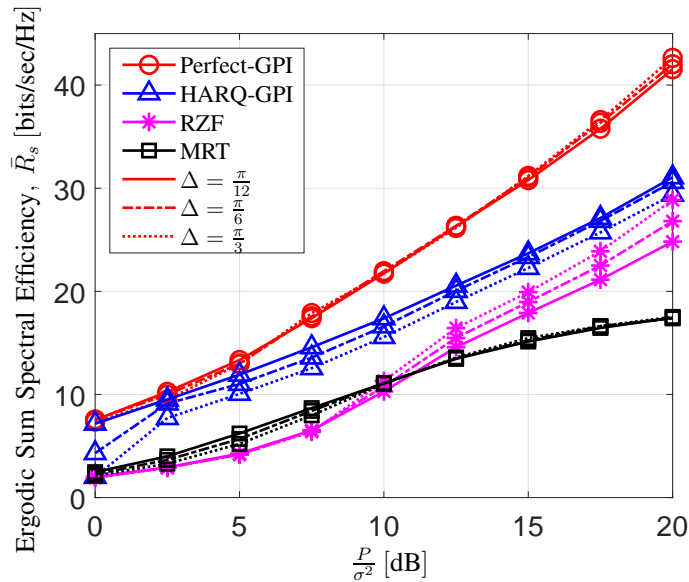}
			\vspace{-10mm}
		}
	\end{center}
	\caption{
		Ergodic sum spectral efficiency $\bar{R}_s$ as a function of $P/\sigma^2$ with $N=8$, $K_t=3$, $K_s=2$, $T=3$, $\delta_4=210$, and $\delta_5=320$ for different values of $\Delta$.
		\vspace{-5mm}
	}
	\label{fig:HARQ_angular}
\end{figure}
%\/\/\/\/\/\/\/\/\/\/\/\/\/\/\/\/\/\/\/\/\/\/\/\/\/\/\/\/\/\/\/\/\/\/\/\/\/\/\/\/\/\/\/\/\/\/\/\/\/\/\/\/\/\/\/\/\/\/\/\/\/\/\/\/\/\/\/\/\/\/\/\/\/\/\/\/\/\/\/\/\/\/\/\/\/\/\/
%

Figure~\ref{fig:HARQ_angular} presents the ergodic sum spectral efficiency $\bar{R}_s$ as a function of $P/\sigma^2$ with $N=8$, $K_t=3$, $K_s=2$, $T=3$, $\delta_4=210$, and $\delta_5=320$ for different values of the angular spread $\Delta$.
Here, we use $\tilde{\gamma}_4(1)=\tilde{\gamma}_4(2)=\tilde{\gamma}_4(3)=1.45$, $\tilde{\gamma}_5(1)=\tilde{\gamma}_5(2)=\tilde{\gamma}_5(3)=1.05$, $w_1=w_2=w_3=1$, and $w_4=w_5=3$.
From this figure,
we can see that as the angular spread $\Delta$ decreases, the difference between the ergodic sum spectral efficiency of the Perfect-GPI and that of the HARQ-GPI reduces.
This is because as $\Delta$ decreases, the inter-beam interference between adjacent antennas decreases.
Therefore, since the approximated spectral efficiencies at the future transmission rounds increase, more transmission power can be allocated to the current transmission round that has well-conditioned channel gain.
However, since the channel correlation between adjacent antennas increases with decreasing $\Delta$, it can be difficult to obtain the time diversity enough (e.g., \ac{RZF} and Perfect-GPI at high $P/\sigma^2$).
From this result, when we do not know the future \ac{CSI} of users, it is advantageous to have a smaller angular spread.

\section{Conclusion} \label{sec:Conclusion}
In this paper, we consider a DCTU-MIMO network, where a \ac{BS} equipped with multiple transmit antennas simultaneously serves delay-constrained users as well as delay-tolerant users.
After analyzing the spectral efficiency of two types of users, we present the lower bound of the spectral efficiency of the delay-constrained user to make it tractable in the optimization.
We then formulate the sum spectral efficiency maximization problem satisfying the latency constraint of the delay-constrained users.
Using the \ac{GPI} precoding algorithm, we propose a computationally efficient algorithm (Delay-GPI) that finds a principal precoding vector that satisfies the first-order optimality condition of the optimization problem.
Furthermore, by dividing a resource frame into multiple time slots, we consider a DCTU-MIMO network with the \ac{IR-HARQ} scheme and propose the HARQ-GPI algorithm to obtain the principal precoding vector.
Finally, we show that the proposed algorithms are better than the baseline schemes.
We also see that the less transmission power is allocated to delay-constrained users, the greater the ergodic sum spectral efficiency is achieved.

%---------------------------------------------------------------------------%
%                                Appendix                                            %
%---------------------------------------------------------------------------%
%
%\begin{appendix}
%	
%\end{appendix}
%
%
%---------------------------------------------------------------------------%
%                                Reference                                            %
%---------------------------------------------------------------------------%

%\bibliographystyle{IEEEtran}
%\bibliography{IEEEabrv,PCP}
%----- [Jemin]: Bibliography
\bibliographystyle{IEEEtran}

\bibliography{StringDefinitions,IEEEabrv,mybib}

\end{document}